\documentclass{nature}
\usepackage{graphicx,psfrag}
\usepackage{mathrsfs}
\usepackage{amsmath,amsfonts,amssymb}
\usepackage{multirow}
\usepackage{comment}
\usepackage{units}
\usepackage{enumerate}
\usepackage[normalem]{ulem} 
\usepackage{longtable}
\usepackage[rgb]{xcolor}
\usepackage{gensymb}
\usepackage{adjustbox}
\usepackage[none]{hyphenat}

\newcommand{\mn}{{Mon. Not. R. Astron. Soc.}}
\newcommand{\mnras}{\mn}
\newcommand{\aj}{{"Astron. J."}}
\newcommand{\apj}{{Astrophys. J.}}
\newcommand{\apjl}{{Astrophys. J. Lett.}}
\newcommand{\apjs}{{Astrophys. J. Supp.}}

\newcommand{\aap}{{Astron. Astrophys.}}
\newcommand{\nat}{{Nature}}

\newcommand{\pasp}{{Pub. Ast. Soc. Pac.}}

\title{A 62-minute orbital period black widow binary in a wide hierarchical triple}

\author{Kevin B. Burdge$^{1,2}$ $^{*}$, Thomas R. Marsh$^{3}$, Jim Fuller$^{4,5}$, Eric C. Bellm$^{6}$, Ilaria Caiazzo$^{4,5}$, Deepto Chakrabarty$^{1,2}$, Michael W. Coughlin$^{7}$, Kishalay De$^{2}$, V.~S. Dhillon$^{8,9}$, Matthew J. Graham$^{4}$, Pablo Rodr\'\i guez-Gil$^{9,10}$, Amruta D. Jaodand$^{4}$, David L. Kaplan$^{11}$, Erin Kara$^{1,2}$, Albert K. H. Kong$^{12}$, S.~R. Kulkarni$^{4}$, Kwan-Lok Li$^{13}$, S. P. Littlefair$^{8}$, Walid A. Majid$^{14,4}$, Przemek Mr{\'o}z$^{15,4}$, Aaron B. Pearlman$^{16,17,4}$, E.~S. Phinney$^{4,5}$, Jan van Roestel$^{4}$, Robert A. Simcoe$^{1,2}$, Igor Andreoni$^{18,19,20}$, Andrew J. Drake$^{4}$, Richard G. Dekany$^{21}$, Dmitry A. Duev$^{4,22}$, Erik C. Kool$^{23}$, Ashish~A.~Mahabal$^{4,24}$, Michael S. Medford$^{25,26}$, Reed Riddle$^{21}$ \& Thomas A. Prince$^{4}$}

\begin{document}

\maketitle

\begin{affiliations}

 \item Department of Physics, Massachusetts Institute of Technology, Cambridge, MA 02139, USA
 \item Kavli Institute for Astrophysics and Space Research, Massachusetts Institute of Technology, Cambridge, MA 02139, USA
 \item Department of Physics, University of Warwick, Coventry CV4 7AL, UK
 \item Division of Physics, Mathematics and Astronomy, California Institute of Technology, Pasadena, CA, USA
 \item TAPIR, Mailcode 350-17, California Institute of Technology, Pasadena, CA 91125, USA
 \item DIRAC Institute, Department of Astronomy, University of Washington, 3910 15th Avenue NE, Seattle, WA 98195, USA
 \item School of Physics and Astronomy, University of Minnesota, Minneapolis, Minnesota 55455, USA
 \item Department of Physics \& Astronomy, University of Sheffield, Sheffield S3 7RH, UK
 \item Instituto de Astrof\'\i sica de Canarias, V\'\i a L\'actea s/n, La Laguna, E-38205 Tenerife, Spain
 \item Departamento de Astrof\'\i sica, Universidad de La Laguna, E-38206 La Laguna, Tenerife, Spain
 \item Department of Physics, University of Wisconsin-Milwaukee, Milwaukee, WI, USA
 \item Institute of Astronomy, National Tsing Hua University, Hsinchu 30013, Taiwan
 \item Department of Physics, National Cheng Kung University, 70101 Tainan, Taiwan
 \item Jet Propulsion Laboratory, California Institute of Technology, Pasadena, CA 91109, USA
 \item Astronomical Observatory, University of Warsaw, Al. Ujazdowskie 4, 00-478 Warszawa, Poland
 \item Department of Physics, McGill University, 3600 rue University, Montreal, QC H3A 2T8, Canada
 \item McGill Space Institute, McGill University, 3550 rue University, Montreal, QC H3A 2A7, Canada
 \item Joint Space-Science Institute, University of Maryland, College Park, MD 20742, USA
 \item Department of Astronomy, University of Maryland, College Park, MD 20742, USA
 \item Astrophysics Science Division, NASA Goddard Space Flight Center, Mail Code 661, Greenbelt, MD 20771, USA
  \item Caltech Optical Observatories, California Institute of Technology, Pasadena, CA 91125, USA
   \item Weights \& Biases, Inc., 1479 Folsom St, San Francisco, CA 94103, USA
 \item The Oskar Klein Centre, Department of Astronomy, Stockholm University, AlbaNova, SE-10691 Stockholm, Sweden
 \item Center for Data Driven Discovery, California Institute of Technology, Pasadena, CA 91125, USA
 \item Department of Astronomy, University of California, Berkeley, Berkeley, CA 94720
 \item Lawrence Berkeley National Laboratory, 1 Cyclotron Rd., Berkeley, CA 94720

\end{affiliations}

\begin{abstract}

Over a dozen millisecond pulsars are ablating low-mass companions in close binary systems. In the original ``black widow'', the 8-hour orbital period eclipsing pulsar PSR J1959+2048 (PSR~B1957+20)\cite{Fruchter1988}, high energy emission originating from the pulsar\cite{Romani2016} is irradiating and may eventually destroy\cite{Phinney1998} a low-mass companion. These systems are not only physical laboratories that reveal the dramatic result of exposing a close companion star to the relativistic energy output of a pulsar, but are also believed to harbour some of the most massive neutron stars\cite{Linares2018}, allowing for robust tests of the neutron star equation of state. Here, we report observations of ZTF J1406+1222, a wide hierarchical triple hosting a 62-minute orbital period black widow candidate whose optical flux varies by a factor of more than 10. ZTF J1406+1222 pushes the boundaries of evolutionary models\cite{Chen2013}, falling below the 80~minute minimum orbital period of hydrogen-rich systems. The wide tertiary companion is a rare low metallicity cool subdwarf star, and the system has a Galactic halo orbit consistent with passing near the Galactic center, making it a probe of formation channels, neutron star kick physics\cite{Igoshev2019}, and binary evolution.
\end{abstract}

Using photometry from the Zwicky Transient Facility\cite{Bellm2019} (ZTF), we searched for short timescale periodic flux variations in 20 million objects that were underluminous relative to the main sequence (Methods) as part of an ongoing campaign to identify short orbital period binary systems\cite{Burdge2020a}. During this search, we identified ZTF J1406+1222, an object which exhibits strong quasi-sinusoidal variability on a period of 62 minutes and a larger amplitude of variability in the ZTF $g$-band than in the ZTF $r$- or $i$-bands.

\begin{figure}
\includegraphics[width=6.5in]{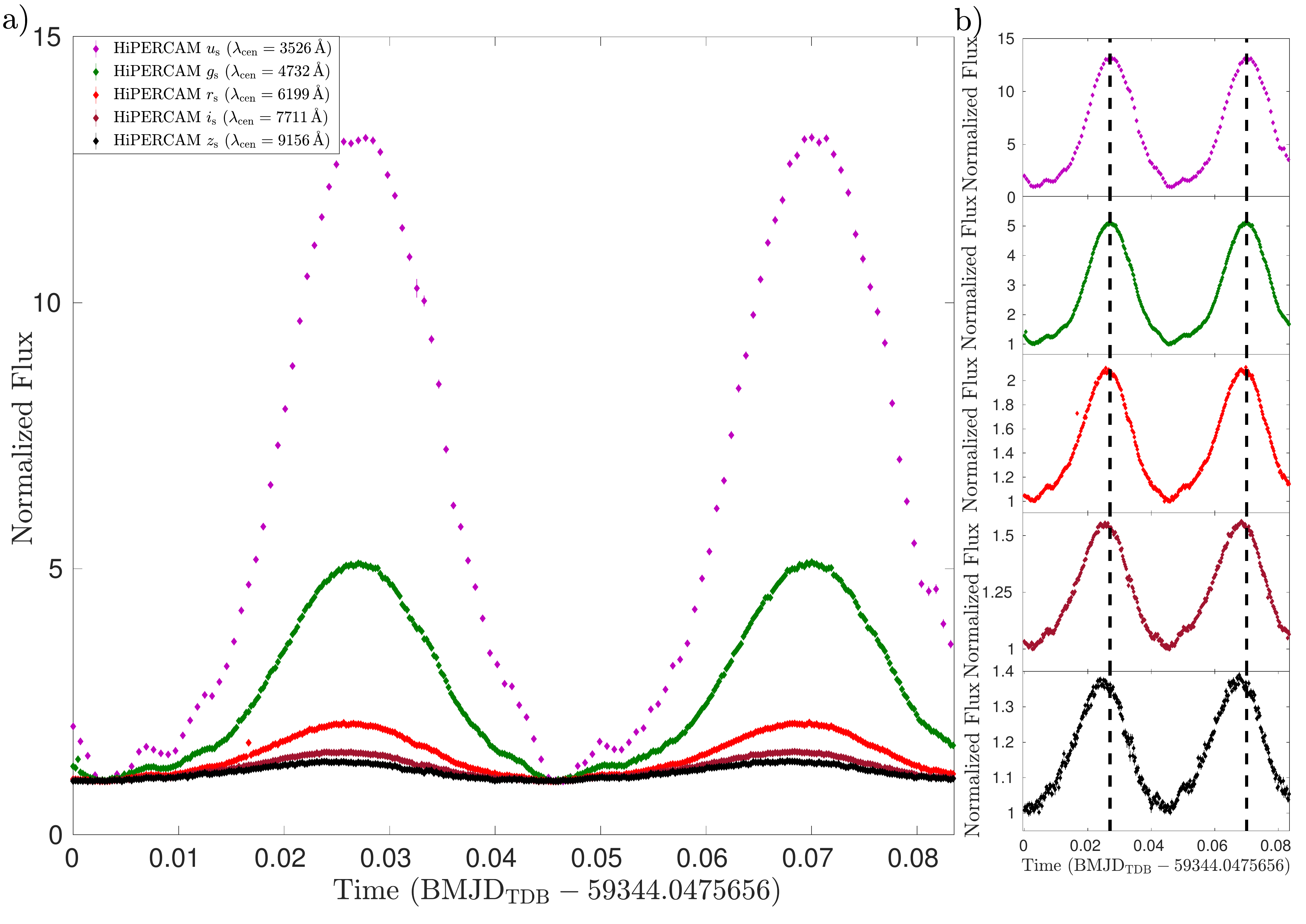}
\linespread{1.0}\selectfont{}
\caption{\textbf{Lightcurve of ZTF~J1406+1222} \textbf{a)} The five-band ($u_{\mathrm{s}},g_{\mathrm{s}},r_{\mathrm{s}},i_{\mathrm{s}},z_{\mathrm{s}}$) HiPERCAM light curve of ZTF J1406+1222. The plotted flux has been normalized relative to the flux level at minimum light in each filter, and the error bars are $1\sigma$ uncertainties.. ZTF J1406+1222 appears to vary significantly more at shorter wavelengths. However, this colour dependence is exaggerated because the tertiary component of the system contributes more in the long wavelength passbands, diluting the apparent amplitude of the signal at longer wavelengths. Because of the contribution of the tertiary component in all optical passbands, the observed degree of brightening, which is already more than a factor of 13 in $u_{\rm s}$, is but a lower bound to the true amplitude of the flux modulation.  \textbf{b)} More detailed plots of the HiPERCAM light curve in each band, with the normalized flux plotted such that the full structure of the variability in each passband is apparent. The vertical dashed black lines indicate the approximate peak flux in $g$, illustrating that the flux at longer wavelengths peaks sooner than at short wavelengths.}

\label{fig:LC}
\end{figure}

\begin{figure}
\includegraphics[width=6.5in]{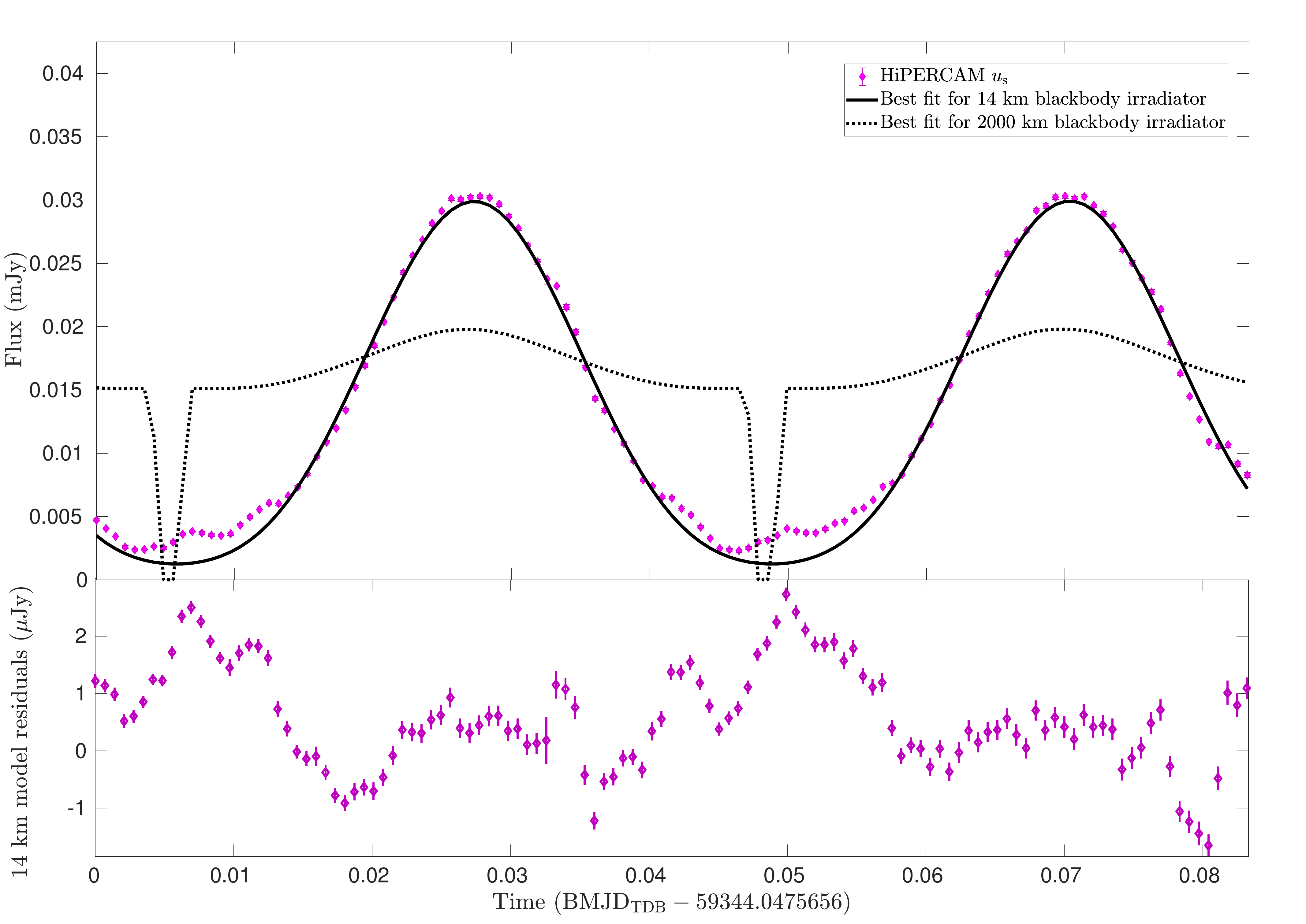}
\linespread{1.0}\selectfont{}
\caption{\textbf{Model fit of the ZTF~J1406+1222 lightcurve} Two model fits to the HiPERCAM $u_{\mathrm{s}}$-band light curve, represented as purple diamonds with error bars representing $1\sigma$ uncertainties. In the model fits, the distance was fixed at 1140~pc. The solid black curve illustrates the best fit using a 14~km blackbody emitter meant to represent a neutron star, whereas the dashed black curve represents the best model fit using a 2000~km blackbody representing a white dwarf as the compact object, which is smaller than the most compact white dwarf known\cite{Caiazzo2021}. The model using a 2000~km blackbody irradiator is unable to reproduce the large amplitude of the light curve because in order to sufficiently increase the amplitude of the irradiation modulation in a system with the observed luminosity at an estimated distance of 1140~pc, the white dwarf temperature must be large ($>80,000\,\rm K$), and in this situation the white dwarf significantly outshines the nightside of the irradiated companion in $u_{\mathrm{s}}$-band, diluting the overall amplitude of variability. }

\label{fig:Model}
\end{figure}

Figure \ref{fig:LC} illustrates a high cadence light curve of ZTF J1406+1222 obtained with the quintuple-beam high speed photometer HiPERCAM\cite{Dhillon2021} on the 10.4-m Gran Telescopio Canarias (GTC) on La Palma. This light curve exhibits a high amplitude and strongly colour-dependent modulation, brightening by more than a factor of 13 in the $u_{\rm s}$-band $(\lambda_{\rm cen}=3526 \rm \, \AA)$. Given the large amplitude, the most plausible physical explanation of its origin is the irradiation of an object by an unseen companion in a 62-minute orbital period binary system: the flux modulation reflects the stark contrast between the ``day'' and ``night'' sides of the irradiated object. Known black widow binaries with identified optical counterparts exhibit this type of strong optical variability, which has been used as a tool to identify several systems near known gamma-ray sources\cite{Kwan-Lok2021,Nieder2020,Clark2021,Pletsch2012,Ray2013}. Light curve models, as seen in Figure \ref{fig:Model}, exclude the possibility of a thermal white dwarf irradiator, as a hot white dwarf would itself outshine the companion, diluting the overall amplitude and resulting in a light curve much like that of the double white dwarf binary ZTF J1539+5027\cite{Burdge2019}. Based on the observed temperature of the heated face of the irradiated object (Methods), we can constrain the heating luminosity, $L_{\mathrm{H}}$ to the range $(1.16 \lesssim L_{\mathrm{H}}\lesssim 1.37)\times10^{34}\,\mathrm{erg}\,\mathrm{s}^{-1}$ (Methods), a value typical of spider binaries\cite{Lee2018}, of which black widows are one sub-class. The modulation peaks earlier at longer wavelengths, suggestive of an asymmetric temperature distribution on the surface of the irradiated object. Such colour-dependent phase shifts in the flux maxima are not seen in irradiated binaries containing white dwarfs, but have been observed in spider binaries, such as PSR J1959+2048 and PSR J2215+5135\cite{Kandel2020}, though the effect is much more pronounced in ZTF J1406+1222, with the $z_{\rm s}$-band flux peaking approximately $10\%$ of an orbital phase earlier than the far ultraviolet Swift UVW2 flux (Methods). The source is not detected in the ultraviolet for over half of its orbit, with a 3-$\sigma$ limit of $>22.9$~\,mag$_\mathrm{AB}$ in the Swift UVW2 filter in an exposure centered on the fainter half of the orbit, further excluding the presence of a hot white dwarf as the irradiating object, which would otherwise have dominated the far ultraviolet flux throughout the orbit.

\begin{figure}
\includegraphics[width=6.5in]{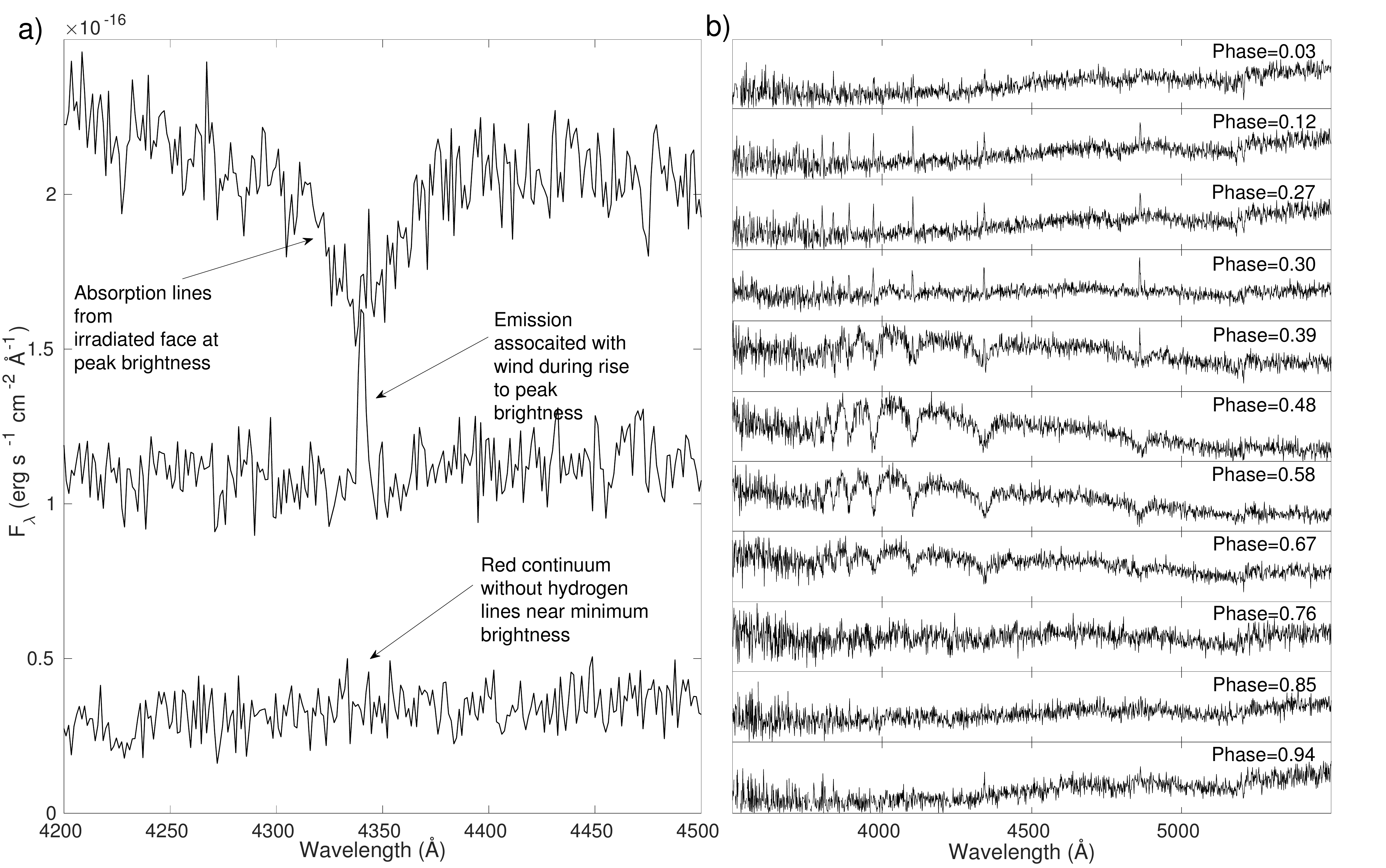}
\linespread{1.0}\selectfont{}
\caption{\textbf{Optical spectroscopy of ZTF~J1406+1222} \textbf{a)} The hydrogen $n=5\rightarrow2$ transition (\rm H$\gamma$) in three phase-resolved spectra of ZTF~J1406+1222 obtained with LRIS on the Keck I telescope. The top spectrum with the largest flux originates from an optically thick phase in which a blue continuum containing hydrogen absorption lines and an appreciable Balmer decrement manifest. In the intermediate flux phase, narrow Balmer emission lines appear, but the continuum is dominated by the tertiary sdK component. In the faintest phase, neither the hydrogen emission nor absorption is visible. \textbf{b)} A collage of eleven 5-minute exposures obtained with LRIS on the Keck (approximately one orbital cycle). Note that the hydrogen absorption lines are visible for approximately four spectra (about 20 minutes, or 1/3 of the 62-minute orbit). The hydrogen emission lines are visible for seven spectra (35 minutes, or over half the orbit), one of which overlaps with a phase that also contains hydrogen absorption. In the bottom spectra near the time of minimum light, no signatures of the hydrogen lines are present, and the only prominent feature remaining is a metal absorption line at $\simeq 5200$~\AA, which is visible throughout the orbit and originates from the sdK tertiary component.}

\label{fig:Spec}
\end{figure}

We obtained phase-resolved spectroscopic observations of the system using the Low Resolution Imaging Spectrometer (LRIS)\cite{Oke1995} on the 10-m W.\ M.\ Keck I Telescope on Mauna Kea. These observations, illustrated in Figure \ref{fig:Spec}, reveal a dramatic transition which occurs on the 62-minute period of the variable. The spectrum is dominated by a red continuum during the faintest phases of the orbital cycle, and then transitions to exhibiting narrow hydrogen emission lines. As the object reaches maximum brightness, a blue continuum with prominent hydrogen absorption lines dominates the flux, but only for $\sim20\%$ of the orbit, before fading back to the faint red continuum. We interpret the appearance of these hydrogen absorption lines as originating from the irradiated dayside of the companion. The high energy radiation which heats the companion in black widow systems penetrates deep enough into the photosphere to produce absorption lines as the reprocessed energy makes its way back to the surface\cite{Romani2015}; this is in stark contrast to systems where the radiation originates from a white dwarf, which is primarily emitted in the ultraviolet, and reprocessed near the surface of the photosphere on the irradiated companion, resulting in an optically thin spectrum dominated by emission lines. The optically thick absorption lines which dominate the spectrum at peak flux are a clear signature of the high-energy irradiation seen in spider binaries.

Deep H$\alpha$ imaging revealed no detectable nebular structure around the object (Methods), and the weak hydrogen emission lines are variable in both intensity and wavelength on the 62-minute period, suggesting that they are not nebular lines as the ones seen in PSR J1959+2048, which are thought to originate primarily from the shocked interstellar medium near the system\cite{Kulkarni1988B}, a large scale structure which extends far beyond the binary. We measured the Doppler shifts of these emission lines and extracted a significant signal, with a semi-amplitude of $112\pm15\,\rm km\, s^{-1}$, confirming the 62-minute orbital period of the system. As seen in the trailed spectra shown in Figure \ref{fig:Trail}, the lines shift bluewards after the time of minimum light, suggesting that they do not originate from the surface of the irradiated object, but rather from an intra-binary shock located between the pulsar and the companion, or in a wind being driven off the irradiated object by the intense pulsar irradiation, forming a cometary tail like that observed in the black widow binary PSR J1311-3430\cite{Romani2015}. The excess luminosity from this feature is likely responsible for the excess flux seen just after minimum brightness in the light curve, as seen in Figure \ref{fig:LC}. 

We did not detect the source in archival data obtained by the Fermi Gamma-ray Telescope\cite{Atwood2009}, in X-ray observations obtained with the Neil Gehrels Swift observatory\cite{Gehrels2004} and the Nuclear Spectroscopic Telescope Array (NuSTAR)\cite{Harrison2013}, nor did we detect the source in radio observations obtained with the 70-m DSS-14 and 34-m DSS-13 antennae at Goldstone Observatory (Methods). Given our estimated distance of $1140$\,pc (Methods), the X-ray sensitivity threshold achieved would have failed to detect several known black widow binaries with similar properties.

In the red portion of the spectrum (Methods) we identified absorption lines characteristic of a rare class of stars known as cool subdwarfs, which are low metallicity, late type stars. Using archival Pan-STARRS1 and SDSS images, \emph{Gaia} eDR3 and the HiPERCAM observations, we confirmed that these originate from a common proper motion wide companion to the inner black widow in a hierarchical triple stellar system (Methods), and its spectroscopic properties are consistent with a late K type cool subdwarf (sdK) (Methods). Given the observed angular separation of $0.5555\pm0.0045$~arcsec between the \emph{Gaia} eDR3 J2016.0 positions of the components and the best distance estimate of 1140~pc, we derive a projected separation of $600$ AU between the components, which corresponds to an orbital period of approximately 12,000~yr, and an estimated projected orbital velocity of just $\approx 1.5$~km\,s$^{-1}$ for the sdK.

A kinematic analysis of ZTF J1406+1222 (Methods) reveals that it is consistent with being a halo object. Hence, its large proper motion of $74.486\pm1.769~\rm mas\,yr^{-1}$ likely reflects this nature, rather than being an indicator of a neutron star natal kick\cite{Lyne1994}. This implies that the system is several billion years old, and thus that the neutron star formed long ago (as its massive progenitor star would have evolved on a timescale of tens of millions of years). It is unlikely that the system could have experienced a significant kick in its current configuration, as it would have easily unbound the tertiary with its estimated orbital velocity of just $\approx \! 1.5$~km\,s$^{-1}$. One intriguing possibility implied by the presence of the wide companion is that the Kozai-Lidov\cite{Kozai1962} mechanism may have played a role in the evolution of the inner black widow\cite{Taylor2017}.

The formation mechanism of ZTF J1406+1222 is an enigma. One plausible origin is that ZTF J1406+1222 was ejected from a globular cluster from a dynamical interaction between two binaries, similar to the formation scenario proposed for PSR J1024{\textendash}0719\cite{Kaplan2016}. Otherwise, if the system originated in the Galactic field, the neutron star must have formed from a low-kick mechanism such as accretion induced collapse (AIC)\cite{Freire2014} (Methods). Although the uncertainties in astrometry and on the age of the system are too large to confidently trace ZTF J1406+1222 to a specific globular cluster, we can determine that it follows an orbit in the Galactic halo which plunges towards the center of the Galaxy, and is consistent with passing less than a fifty parsecs from the Galactic Center (Methods). It is thought that globular clusters interacting with the Galactic center can be disrupted, and that the gamma-ray excess near the Galactic center originates from millisecond pulsars left behind by these disrupted clusters\cite{Fragione2018}. The average density of a globular cluster is on the order of $10^3--10^5\,M_{\odot}\,\rm pc^{-3}$, much lower than the average density of matter within the triple ZTF J1406+1222 ($\sim10^8\,M_{\odot}\,\rm pc^{-3}$), meaning that an interaction with the center of the Galaxy which disrupted the globular cluster would likely not unbind ZTF J1406+1222, allowing it to instead be ejected in its current state. This suggests that ZTF J1406+1222 could have originated from a millisecond pulsar rich globular cluster which was disrupted near the Galactic center. 

The hierarchical triple nature and short orbital period of ZTF J1406+1222 distinguish it from any known spider binary, challenging formation models of these systems. Uniquely among spider binaries,  ZTF J1406+1222 was identified using only optical photons, a novel method of discovering binaries hosting neutron stars solely on the basis of their strongly irradiated companions, thus potentially eliminating strong selection effects in previous work, which identified them primarily on the basis of their emission at radio, X-ray, or gamma-ray wavelengths.

\begin{figure}
\includegraphics[width=6.5in]{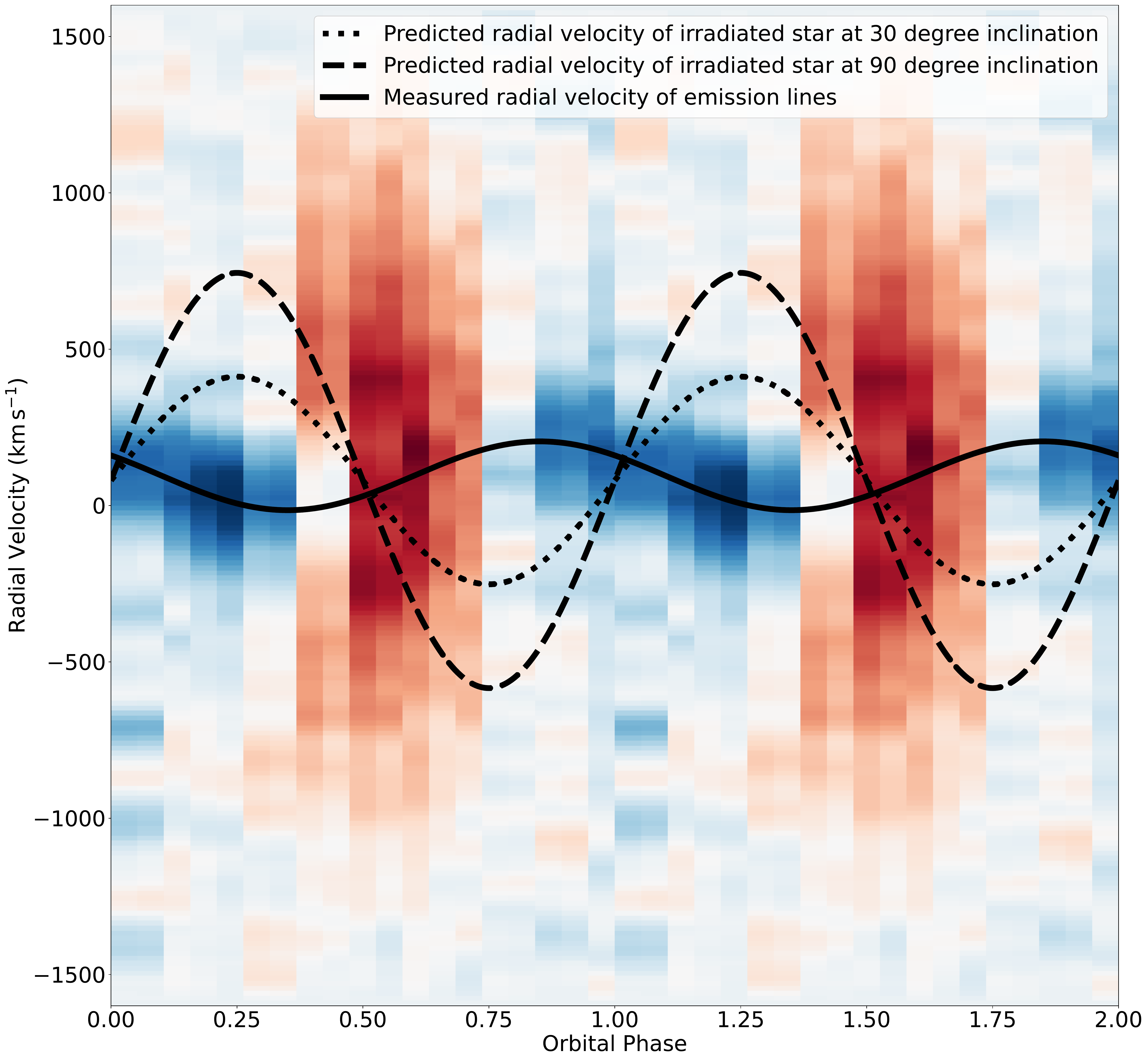}
\linespread{1.0}\selectfont{}
\caption{\textbf{ZTF~J1406+1222's trailed spectra} The trailed LRIS spectra of ZTF J1406+1222 illustrating the variations in brightness of a sum of the hydrogen lines over the course of an orbit (with red representing absorption, and blue emission), projected onto velocity space on the y-axis. The orbital phase has been referenced to the time of maximum light in the HiPERCAM $u_{\rm s}$ lightcurve (which corresponds to phase $0.5$). The blue central features are the hydrogen emission lines visible in the spectrum, while the red broader lines which appear near phase 0.5 are the absorption lines associated with the irradiated companion. A velocity modulation on the 62-minute period is clearly apparent in the emission lines, with an estimated semi-amplitude of $112\pm15\,\rm km\, s^{-1}$, confirming that this is the orbital period. The solid black line is the best sine fit to the emission-line velocities, which are associated likely associated with the intrabinary shock in the system, or a wind driven off of the companion. The dashed black lines indicate the expected radial velocity variation of the irradiated star in the binary, assuming $\sim 30$-degree and $\sim 90$-degree inclinations and a $0.05\,\mathrm{M}_\odot$ object orbiting a $1.4\,\mathrm{M}_\odot$ unseen companion.}

\label{fig:Trail}
\end{figure}

\begin{table}
\renewcommand{\thetable}{\arabic{table}}
\centering
\caption{Table of parameters} \label{tab:Parameters}
\begin{tabular}{cc}

\hline
\hline
$RA\,\,\rm (BW)$ &$211.7340580\pm	1.22\times10^{-6}\,\mathrm{deg}\,\,\rm (Epoch\,\,J2016.0,\,\, Equinox\,\,J2000)$ \\
   \hline
   $Dec\,\,\rm (BW)$ &$12.37872188\pm9.48\times10^{-7}\,\mathrm{deg}\,\,\rm (Epoch\,\,J2016.0,\,\, Equinox\,\,J2000)$ \\
   \hline
   $PM\,RA$ &$-73.824\pm1.135\,\mathrm{mas}\,\,\rm (Epoch\,\,J2016.0,\,\,Equinox\,\,J2000)$\\
   \hline
   $PM\,Dec$ &$-9.912\pm1.137\,\mathrm{mas}\,\,\rm(Epoch\,\,J2016.0,\,\,Equinox\,\,J2000)$\\
   \hline
   $D$ &$1140\pm200\,\mathrm{pc}$\\
   \hline
   $\gamma$ &$67\pm30\,\mathrm{km\,s^{-1}}$\\   
   \hline
   \hline
   $  T_\mathrm{dayside}$ & $10462\pm150\,\mathrm{K}$  \\
   \hline
   $  T_\mathrm{nightside}$ & $<6500\,\mathrm{K}$  \\
   \hline
   $\rho_{\mathrm{Irradiated}}$& $>10\,\rm g \, cm^{-3}$ \\
   \hline
   $R_{\mathrm{Irradiated}}$& $>0.03\,R_{\odot}$ \\
   \hline
   $L_{\mathrm{H}}$ & $(1.13 <L_{H}< 1.79)\times10^{34}\,\mathrm{erg}\,\mathrm{s}^{-1}$  \\
   \hline
   $  L_\mathrm{x}$ & $<2.9\times10^{30},\mathrm{erg}\,\mathrm{s}^{-1}\,\rm (at\,\,1140\,pc)$  \\
   \hline
   $T_{0} (\mathrm{in\,\,} u_{\mathrm{s}})$ & $59344.053345\pm0.000039\,   \mathrm{BMJD_{TDB}}$  \\
   \hline
   $P_{\mathrm{b}}$ & $3720.09213\pm0.00021\,  s$  \\
   \hline
   \hline
  $RA\,\,\rm (sdK)$ &$211.7342086\pm	2.27\times10^{-6}\,\mathrm{deg}\,\,\rm (Epoch\,\,J2016.0,\,\, Equinox\,\,J2000)$ \\
   \hline
   $Dec\,\,\rm (sdK)$ &$12.37860287\pm2.66\times10^{-7}\,\mathrm{deg}\,\,\rm (Epoch\,\,J2016.0,\,\, Equinox\,\,J2000)$ \\
  \hline
   $P_{b} (\mathrm{sdK})$ & $>10000\,\mathrm{yr}\,\rm (at\,\,1140\,pc)$  \\
   \hline
    $T_\mathrm{eff} (\mathrm{sdK})$&$4020\pm70 \rm \, K$\\
\end{tabular}
\end{table}

\newpage


\begin{addendum}
 \item K.B.B. is a postdoctoral fellow in the MIT Pappalardo Fellowships in Physics, and thanks the program, the MIT Physics Department, and the MIT Kavli Center for Space Research, for ongoing support of his research. T.R.M. was supported by a Leverhulme Research Fellowship and STFC grant ST/T000406/1. I.C. is a Sherman Fairchild Fellow at Caltech and thanks the Burke Institute at Caltech for supporting her research. M.W.C acknowledges support from the National Science Foundation with grant number PHY-2010970. A.B.P. is a McGill Space Institute~(MSI) Fellow and a Fonds de Recherche du Quebec -- Nature et Technologies~(FRQNT) postdoctoral fellow. The design and construction of HiPERCAM was funded by the European Research Council under the European Union’s Seventh Framework Programme (FP/2007-2013) under ERC-2013-ADG Grant Agreement no. 340040 (HiPERCAM). VSD and HiPERCAM operations are supported by STFC grant ST/V000853/1. E.C.K. acknowledges support from the G.R.E.A.T research environment funded by {\em Vetenskapsr\aa det}, the Swedish Research Council, under project number 2016-06012, and support from The Wenner-Gren Foundations.
 J.F. and E.S.P. acknowledge support from the Gordon and Betty Moore Foundation
through Grant GBMF5076 to E.S.P.
 E.C.B. acknowledges support from the NSF  AAG  grant  1812779  and  grant  \#2018-0908 from the Heising-Simons Foundation.
 K.L.L. is supported by the Ministry of Science and Technology of the Republic of China (Taiwan) through grant 110-2636-M-006-013, and he is a Yushan (Young) Scholar of the Ministry of Education of the Republic of China (Taiwan).
 
 Based on observations obtained with the Samuel Oschin Telescope 48-inch and the 60-inch Telescope at the Palomar Observatory as part of the Zwicky Transient Facility project. ZTF is supported by the National Science Foundation under Grant No. AST-1440341 and a collaboration including Caltech, IPAC, the Weizmann Institute for Science, the Oskar Klein Center at Stockholm University, the University of Maryland, the University of Washington, Deutsches Elektronen-Synchrotron and Humboldt University, Los Alamos National Laboratories, the TANGO Consortium of Taiwan, the University of Wisconsin at Milwaukee, and Lawrence Berkeley National Laboratories. Operations are conducted by COO, IPAC, and UW.
 
Some of the data presented herein were obtained at the W.M. Keck Observatory, which is operated as a scientific partnership among the California Institute of Technology, the University of California and the National Aeronautics and Space Administration. The Observatory was made possible by the generous financial support of the W.M. Keck Foundation. The authors wish to recognize and acknowledge the very significant cultural role and reverence that the summit of Mauna Kea has always had within the indigenous Hawaiian community. We are most fortunate to have the opportunity to conduct observations from this mountain.
 
 \item[Competing Interests] The authors declare that they have no
competing financial interests.
 \item[Contributions] KBB discovered the object, conducted the LCURVE and ICARUS light curve analysis, spectroscopic data reduction and analysis of Keck LRIS data, the reduction and analysis of the Swift UVOT and XRT data, reduction and analysis of the NuSTAR data, and was the primary author of the manuscript. TRM conducted the HiPERCAM data reduction and assisted with the overall analysis and interpretation of the object. KBB performed the analysis of the Fermi LAT data, with assistance from DLK, AKHK and KLL. WAM and ABP conducted the DSN observations of the system and ABP performed the pulsation searches of these data and contributed the text describing the results and the implications. JF contributed text regarding the evolutionary history of the system. TAP, TRM, JF, IC and ESP all contributed to the interpretation of the object's evolutionary history. All authors contributed comments and edits to the manuscript. ADJ submitted the NuSTAR proposal on the object and KD conducted the WASP observations of the object. VSD conducted the HiPERCAM observations of the object and is PI of HiPERCAM. PRG was PI of the HiPERCAM proposal which observed the object. 
SRK, TAP, MJG, and ECB are respectively the PI, Co-I, project scientist, and survey scientist of ZTF.
 \item[Correspondence] Correspondence and requests for materials
should be addressed to K.B.B.~(email: kburdge@mit.edu).
\end{addendum}

\newpage

\begin{methods}

\subsection{Discovery and confirmation}

We searched a sample of objects selected as being potentially under-luminous relative to the main sequence using \emph{Gaia} Data Release 2 (DR2) astrometry. The selection is described by
\begin{equation}
    \label{eq:selection}
   G_\mathrm{mean}+5\times\left[\log(\sigma+\sigma_\mathrm{err})-2\right]>5\times (BP-RP)+8~,
\end{equation} where $G_\mathrm{mean}$ is the mean \emph {Gaia} G apparent magnitude and $(BP-RP)$ is the colour, both from Gaia DR2. This selection yielded a total of 22,912,186 candidate objects, which we then cross-matched with data from the Zwicky Transient Facility\cite{Bellm2019,Graham2019,Dekany2020,Masci2019}. We searched for periodic signals in all objects which had more than 50 combined epochs in their $g$ and $r$ light curves using the conditional entropy algorithm\cite{Graham2013} and selected candidates whose minimum conditional entropy fell more than 20 standard deviations below the mean entropy of the power spectrum. We searched frequencies ranging from $720>f>\frac{2}{\rm baseline}\rm\,cycle\,day^{-1}$, where the baseline is the time elapsed between the first and last epochs in each light curve in days. This selection yielded 25,025 high significance candidate objects, whose phase-folded light curves were visually inspected (investigation of many of these candidates is ongoing). ZTF J1406+1222 immediately captured our attention due to the stark difference in modulation amplitude between its ZTF $g$, ZTF $r$ and ZTF $i$ light curves. In particular, it exhibited a larger amplitude at blue wavelengths, as seen in Extended Data Figure 1. This is unexpected for white dwarf binaries with irradiated companions at such short orbital periods, as they host cool late type stars, which preferentially reprocess radiation at longer wavelengths. We thus immediately suspected that ZTF J1406+1222 could be a black widow in which the companion was so heavily irradiated that its inner face had heated sufficiently to preferentially reprocess radiation in bluer wavelengths. However, we later discovered that this difference in amplitude as a function of wavelength is primarily due to the presence of the tertiary sdK companion which was diluting the apparent amplitude at redder wavelengths.

\begin{figure}
\includegraphics[width=6.5in]{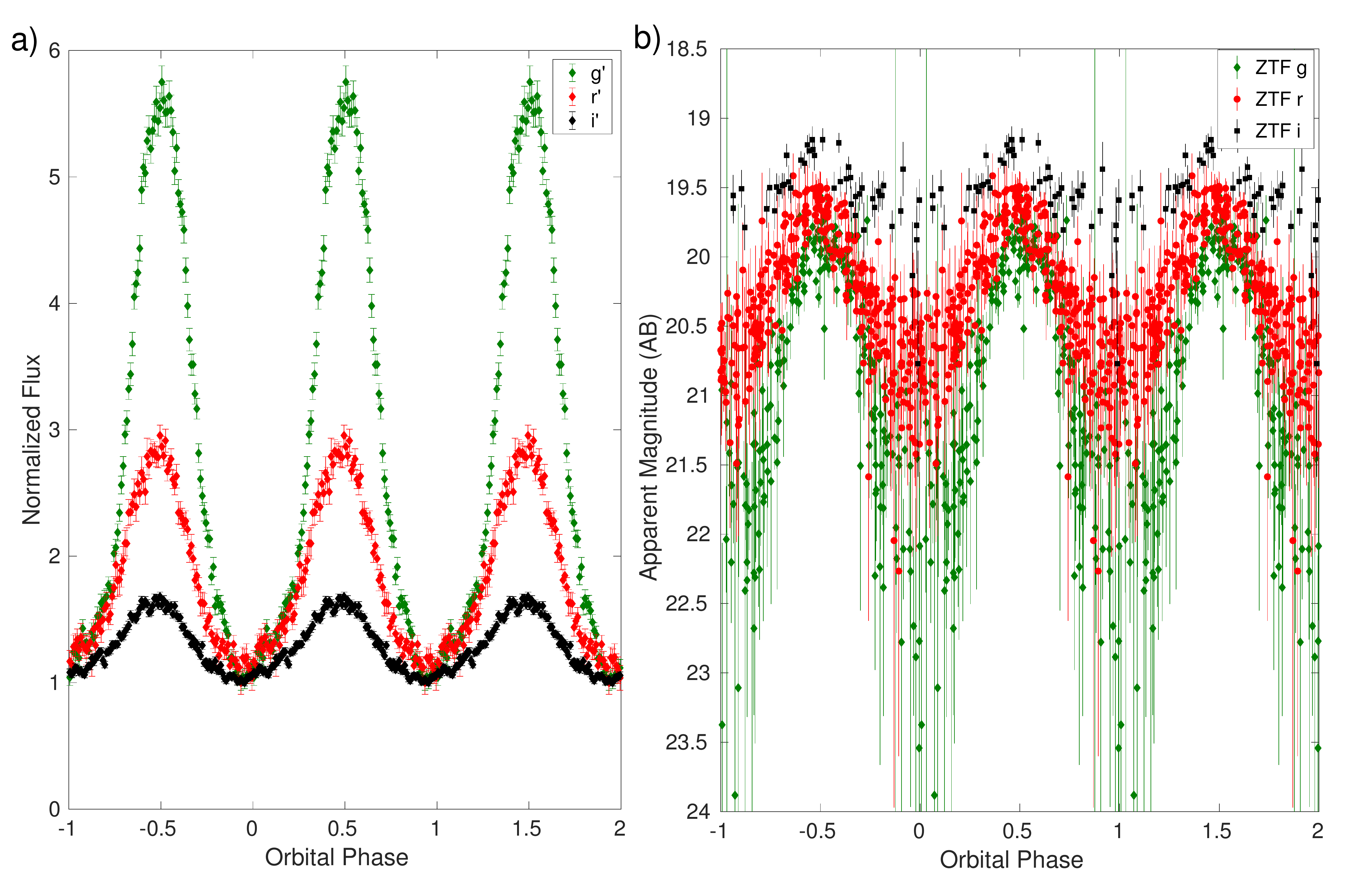}
\linespread{1.3}\selectfont{}
\renewcommand{\figurename}{Extended Data Figure}
\setcounter{figure}{0}  
\caption{\textbf{CHIMERA and ZTF lightcurve of ZTF J1406+1222} \textbf{a)} The CHIMERA light curve of ZTF J1406+1222. Like the HiPERCAM light curve illustrated in Figure \ref{fig:LC}, the variability is much more pronounced at shorter wavelengths. This light curve was used to confirm the signal observed in the ZTF light curve, prompting further follow-up of the target. \textbf{b)} The ZTF light curve of ZTF J1406+1222. The object appears more variable in the shorter wavelength passbands, but is brighter overall at longer wavelengths, primarily due to the contribution of the sdK tertiary component, and thus the true fractional variability is underestimated in redder passbands. All the error bars illustrated represent $1\sigma$ flux uncertainties.}

\label{fig:LC2}
\end{figure}

Using the high-speed imager CHIMERA\cite{Harding2016} on the 200-inch Hale telescope at Palomar Observatory on the nights of July 15 and 21 2020, we obtained the $g^{\prime}$, $r^{\prime}$, and $i^{\prime}$ light curves illustrated in Extended Data Figure 1, confirming the 62-minute period modulation of over a factor of 5 in $g^{\prime}$ and smaller amplitudes in $r^{\prime}$ and $i^{\prime}$. We reduced these data using a custom pipeline which performed aperture photometry on the source and a nearby reference star in each frame using a dynamic full-width-half-max.

We also used the Wafer Scale Imager at Prime (WASP) to obtain a 1200-s an H$\alpha$ filter ($\lambda_{0}=6570\,\mathrm{\AA},\Delta \lambda=100\,\mathrm{\AA}$) and an additional 1200-s exposure in an off-band filter ($\lambda_{0}=6651\,\mathrm{\AA},\Delta \lambda=107\,\mathrm{\AA}$) to check for the presence of a nebula around the object. We bias subtracted and flat fielded these observations with a custom pipeline. We then subtracted the normalized off-band exposure from the normalized H$\alpha$ exposure and found no discernible extended emission in the resulting image.  

Data using the three channel imager ULTRACAM\cite{Dhillon2007} on the 3.5-m New Technology Telescope at LaSilla, taken in $u$, $g$ and $i$ filters, were used to confirm that the amplitude of the light curve modulation increased substantially in the $u$ filter relative to what we had observed previously in redder filters.

After confirming the extreme amplitude of modulation in the $u$ filter, we obtained two hours of data using the quintuple beam high speed photometer HiPERCAM\cite{Dhillon2021} on the 10.4-m Gran Telescopio Canarias and used these data for our light curve modelling.

\subsection{HiPERCAM light curve analysis}

Modelling the HiPERCAM light curve of ZTF J1406+1222 is challenging for three reasons: 1. We lack kinematic constraints and thus the binary mass ratio and the scale of the semi-major axis are not constrained; 2. the two modelling codes we used, LCURVE\cite{Copperwheat2010} and ICARUS\cite{Breton2012}, lack the physics needed to fully describe the behaviour we see in the light curve, in particular, contributions from the wind near minimum flux and the phase shift seen at maximum; and 3. the cool subdwarf companion contributes some amount of light in each band and its assumed spectral energy distribution greatly impacts the amplitude of the modulation, especially in the redder filters. We introduce a correction for this contribution in our ICARUS model. 

Despite these limitations, we can still make useful constraints on the nature of the system using light curve models by investigating simple limiting cases. As discussed in the section on distance estimates, we know the distance to the system must be $D<2$\,kpc based on its large proper motion, we are able to precisely measure its apparent magnitude at peak and minimum at wavelengths ranging from the ultraviolet to near infrared, we know that there is little extinction along the line of sight, and thus we can estimate the peak luminosity of the system as a function of wavelength and also get an upper limit on the minimum luminosity of the irradiated object (the wind contributes near minimum flux, so the true depth of the minimum is unknown). We also know the orbital period of the system with a high degree of precision and can therefore assume a total system mass and estimate the scale of its semi-major axis. 

We developed a model based on the HiPERCAM $u_{\rm s}$-band light curve due to the minimal contamination from the sdK companion at these wavelengths and high signal-to-noise ratio. We used LCURVE to model the light curve and performed nested sampling using the PyMultiNest \cite{Buchner2014,Feroz2009} package. For this model, we ignored the third light contribution due to the subdwarf, which is small in the $u_{\rm s}$ band.

The purpose of our LCURVE model was to explore whether a model using a neutron star and/or a white dwarf irradiator could acceptably fit the data. Due to the clear contribution of the wind near flux minimum, we adjusted the weights of all but the most restrictive point near the minimum to 0 (even this point likely has some wind contribution). We fixed the orbital period at $P=3720.09213$\,s ,based on the ZTF light curve, and the mass ratio at $q = 0.0357$ (assuming a mass of $0.05\,\mathrm{M}_\odot$ for the irradiated object and $1.4\,\mathrm{M}_\odot$ for the compact object). We adopted limb-darkening coefficients for a model atmosphere of a $10\,000\,$K main sequence star (we use this value to reflect the temperature of the irradiated surface) with a surface gravity $\log(g)=5.5$\,dex (cgs units). The mass ratio was fixed because we found that when allowed to vary freely, the solution strongly favoured irradiated objects of arbitrarily high mass in order to increase the radius of the irradiated object without filling its Roche lobe. We then directly fit the $u_{\rm s}$-band flux level of the light curve by introducing the distance as a parameter in the model (extinction is negligible along this line of sight, with a reddening of $\rm E(g-r)<0.02$ within a distance of $6\,\rm kpc$\cite{Green2019}). We constructed two models using the method described here, with the radius of the compact object responsible for emitting the irradiating flux fixed to 14\,km or 2000\,km, respectively. We found that the latter model was not even remotely able to fit the data, as seen by the fit presented in Figure~\ref{fig:Model}. As a check, we allowed the distance to vary within the 2000~km model and observed that LCURVE was able to converge to an acceptable solution for an object approximately 6.6\,kpc away by increasing the nightside temperature of the irradiated object to significantly outshine the extremely hot 2000~km irradiator even at flux minimum. The model is not able to achieve this at smaller distances, as making the irradiated object's night side this bright increases the overall luminosity of the system too much to be consistent with the observed apparent magnitude. Increasing the assumed mass of the neutron star in the model above $1.4\,\mathrm{M}_\odot$ results in a larger binary semi-major axis, and increases the total required heating flux radiated by the neutron star.

These simplified LCURVE models were meant to test the possibility of a white dwarf compact object. However, LCURVE treats the objects in the system as simple black bodies and neglects atmospheric corrections such as the attenuation in flux seen in the $u_{\rm s}$ band relative to a simple blackbody due to the Balmer ionization. Therefore, to construct the most realistic model possible now that we had ruled out the possibility of a white dwarf solution, we turned to the ICARUS light curve modelling code. 

Before constructing the ICARUS model, we corrected for the contribution of the sdK component. The flux-weighted
position vector in the HiPERCAM images of the combined black widow plus sdK star can be written as
\begin{equation}
    \mathbf{r} = \frac{f_{\rm bw} \mathbf{r}_{\rm bw} + f_{\rm sd} \mathbf{r}_{\rm sd}}{f_{\rm bw} + f_{\rm sd}}~.
\end{equation}
The flux of the sdK, $f_{\rm sd}$, is constant, and although the positions of the two stars move due to telescope guiding errors, there should be a fixed offset $\mathbf{s}$ between them, i.e.
\begin{equation}
    \mathbf{r}_{\rm sd} = \mathbf{r}_{\rm bw} + \mathbf{s}~.
\end{equation}
We cannot measure the individual fluxes because of the blending, and instead measure the total
\begin{equation}
    f = f_{\rm sd} + f_{\rm bw}~.
\end{equation}
Using these relations to substitute for $f_{\rm bw}$ and $\mathbf{r}_{\rm sd}$, the position of the centroid can be written as
\begin{equation}
    \mathbf{r} = \mathbf{r}_{\rm bw} + \frac{f_{\rm sd}}{f}\, \mathbf{s}~.
\end{equation}
The position vector of the black widow in this expression is still subject to guiding errors, so we reference the position with respect to the comparison stars to remove the variability:
\begin{equation}
    \Delta\mathbf{r} = \Delta \mathbf{r}_{\rm bw} + (f_{\rm sd}\,\mathbf{s}) (1/f)~,
\end{equation}
where now $\Delta\mathbf{r}_{\rm bw}$ is now a constant difference relative to the comparison star.
This equation shows that the centroid position of the blended image varies linearly with the inverse of the flux, $1/f$. If the measurements are made along a line parallel to $\mathbf{s}$, then a plot of position versus $1/f$ has gradient $f_{\rm sd}\, \mathbf{s}$. We know $\mathbf{s} = 0.5555$\,arcsec from Gaia, hence the contribution of the sdK to the flux, $f_{\rm sd}$, results. This has to be applied to the data for each filter, and the resultant corrected lightcurves are plotted in Extended Data Figure 2.

In ICARUS, we fixed the mass ratio $q=0.0357$ (again assuming a mass of $0.05\,\mathrm{M}_\odot$ for the irradiated object and $1.4\,\mathrm{M}_\odot$ for the compact object), the orbital period $P=3720.09213$\,s and the distance $d=1140$\,pc. ICARUS is designed to fit multi-band light curves simultaneously. However, because of the significant wavelength dependent phase shifts in the data, ICARUS was unable to converge to an acceptable solution by simply fitting the light curves without correcting for these shifts. We arbitrarily shifted the light curves to have all other maxima at the $u_{\rm s}$-band peak flux phase and performed a series of ICARUS fits. In order to account for the wind contribution, we experimented with not including data near the minimum flux in the fit. We found that because of the extraordinarily high signal-to-noise ratio of the HiPERCAM data, Markov-chain Monte Carlo (MCMC) fits of models which accounted for only data between phases $0.4-0.6$ (where $0.5$ is the flux maximum), vs $0.3-0.7$, vs all of the data converged to radically different solutions, with orbital inclinations in the $35-90$\,deg range. In fact, without any constrains, there were even good fits achieved for inclinations as low as $14$ degrees, but these solutions were unphysical, as they required a mass ratio in which the irradiated object was far more massive than the accretor, and that the system be at a great distance, as the solutions were trying to force the radius of the irradiated object to be as large as possible. Because the ICARUS model does not account for physics such as flux from the intrabinary shock (which may have an asymmetric offset from the line between the pulsar and the companion), the wind being driven off of the companion or the possibility of atmospheric transport on the surface of the irradiated object, rather than derive parameters from these models, we choose not to report these parameter estimates. A future detection of the pulsar, which would provide a precisely measured time of superior conjunction and an estimate of its radial velocity, may provide enough additional constraints for a more sophisticated modelling of the system and avoid simply over-fitting the data. Our ICARUS and LCURVE models both generally preferred solutions in which the irradiated object was underfilling its Roche lobe (with a $<0.5$ filling factor); however, the inferred filling factor was sensitive to which portion of the data was modelled, and thus is poorly constrained. An example ICARUS model fit is illustrated in Extended Data Figure 3.

However, in addition to unambiguously ruling out the possibility of a white dwarf irradiator, the light curve also allows us to estimate the amount of radiation the neutron star is emitting, $\dot{E}$. This is because the peak temperature of the irradiated companion, $T_{\rm day}$, can be estimated from the apparent magnitudes at peak flux, and this temperature is directly related to the amount of incident radiation per unit area by the follow expression:
\begin{equation}
    \label{ddot}
    L_\mathrm{H}\sim4\pi a^{2}\sigma(T_{\mathrm{day}}^4-T_{\mathrm{night}}^4)~,
\end{equation}where $L_\mathrm{H}$ is the heating luminosity, $\sigma$ is the Stefan-Boltzmann constant, $T_{\rm night}$ is the nightside temperature of the companion and $a$ is the semi-major axis of the binary system. The heating luminosity and the spin down luminosity are related by an efficiency factor, $L_\mathrm{H}=\eta \dot{E}$. Although we do not know the total mass of the system, $M_\mathrm{T}$, and therefore must assume one to estimate the semi-major axis, $a$ only varies as $M_\mathrm{T}^{1/3}$ and is much more sensitive to the orbital period, which we know quite precisely. We consider a range of semi-major axis values, assuming $1.45<M_\mathrm{T}<2.05\,\mathrm{M}_\odot$. We estimate $T_{\rm day}$ by fitting the spectral energy distribution (SED) at the orbital phase of maximum flux in the Swift UVOT light curve, and our best fit model is illustrated in Extended Data Figure 4. This gives a warmer estimate than if we simply take the peak flux for each filter, since redder filters peak earlier than blue filters. The companion, given the short orbital period, must be degenerate or near degenerate, and its atmosphere is clearly dominated by hydrogen lines. Thus, in order to fit the SED of the dayside of the object, we used hydrogen rich low-mass white dwarf atmospheric models\cite{Gianninas2014,Tremblay2015}, which yielded an estimated temperature of $T_\mathrm{day}=10462\pm150$\,K. To obtain the most conservative estimate of $T_\mathrm{night}$, we constructed an ICARUS model in which we fix $T_\mathrm{day}$ to $10462\pm150\,\mathrm{K}$ and fix the inclination to $90\,\mathrm{deg}$. This requires the light curve model to pick the largest possible value of $T_\mathrm{night}$ in order to minimize the amplitude of variability to match the data. In this limiting case, find $T_\mathrm{night}\approx6500\,\mathrm{K}$. Using these boundaries and assuming a semi-major axis corresponding to a total system mass of $1.45<M_\mathrm{T}<2.05\,\mathrm{M}_\odot$, we conclude that $L_{\mathrm{H}}$ is approximately in the range of $(1.13 <L_{\mathrm{H}}< 1.79)\times10^{34}\,\mathrm{erg}\,\mathrm{s}^{-1}$, a value typical of spider binaries.

\begin{figure}
\includegraphics[width=6.5in]{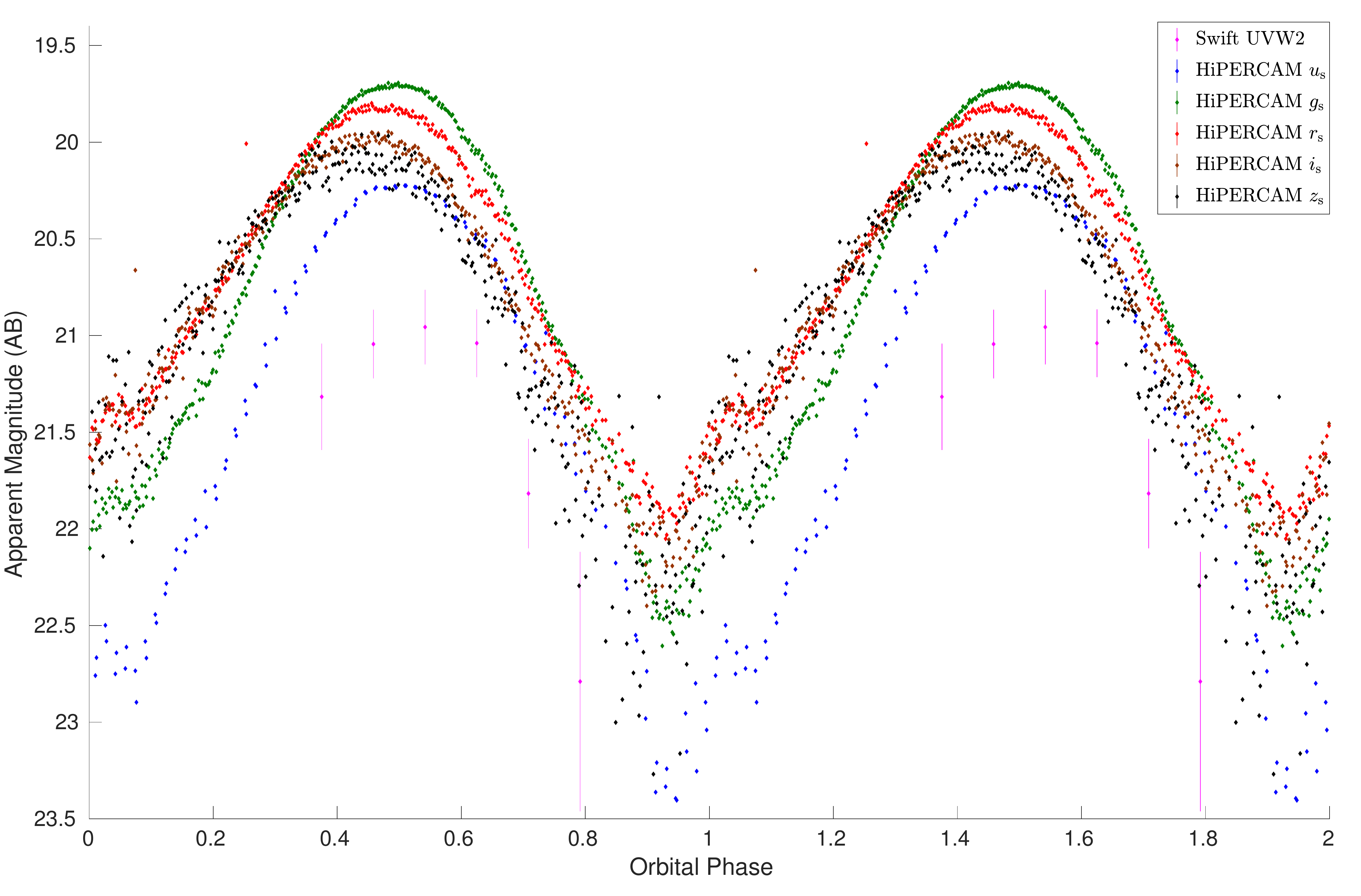}
\linespread{1.3}\selectfont{}

\renewcommand{\figurename}{Extended Data Figure}
\caption{\textbf{Swift UVOT and corrected HiPERCAM lightcurve of ZTF J1406+1222} The HiPERCAM and Swift UVOT light curves on a magnitude scale after correcting for the contribution of the sdK star. Only $>1\sigma$ detections have been shown for the Swift data, as the object magnitude lies below the detection threshold for over half of the orbit. The prominent phase shifts in the peak flux indicate that the object transitions from a cooler to a hotter surface over the course of the flux maximum. Our LCURVE and ICARUS models were not able to account for this physical effect, so we arbitrarily shifted the other light curves to match the $u_{\rm s}$-band maximum flux for the purpose of constructing simple models. In order to estimate the temperature of the day side of the irradiated object, we fitted the spectral energy distribution from the apparent magnitudes at orbital phase 0.55, the peak of the Swift UVOT UVW2 light curve. One orbital cycle has been repeated for display purposes. All the error bars illustrated represent $1\sigma$ uncertainties in apparent magnitude.}

\label{fig:Mags}
\end{figure}

\begin{figure}
\includegraphics[width=6.5in]{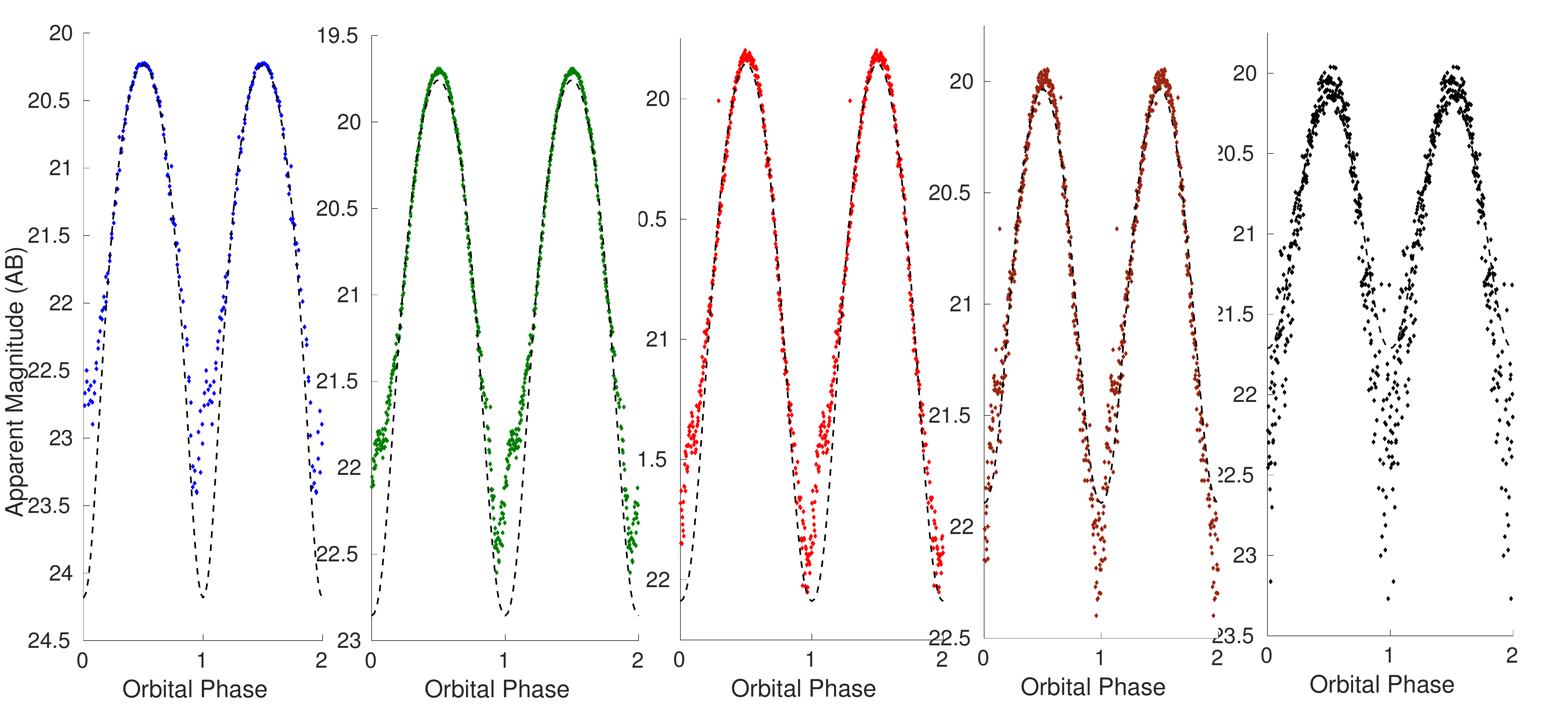}
\linespread{1.3}\selectfont{}
\renewcommand{\figurename}{Extended Data Figure}
\caption{\textbf{ICARUS model fit to lightcurve} An example ICARUS toy model fit to data between orbital phases 0.25 and 0.75 (ignoring data outside these phases), with the temperature of the day side of the companion fixed to $10462\,\mathrm{K}$, the distance fixed to $1140\,\mathrm{pc}$, a fitted inclination of $i=\approx66\,\mathrm{deg}$ and an irradiated object with a radius of just $0.029\,\mathrm{R}_\odot$. The data from left to right illustrate the HiPERCAM $u_{\rm s}$, $g_{\rm s}$, $r_{\rm s}$, $i_{\rm s}$, and $z_{\rm s}$ filters, with the dashed black lines illustrating the best fit model in that filter. The light curves have been artificially shifted to line up with the $u_{\rm s}$-band maximum light, as the ICARUS model is unable to capture the strong colour-dependent phase shifts seen in the data. We did not perform an MCMC over these models, as we found that there were acceptable fits for a wide range of inclinations and other parameters given the limitations of using a simple direct heating model. Because we chose to fit the flux at peak, but not at minimum (where the wind contributes significantly), most models fit the peak flux well, but the excess flux at minimum increases gradually to the redder passbands, suggesting that there may be a colour dependence associated with the flux contribution from the wind, though this is difficult to disentangle from the flux contribution of the sdK at minimum light. }

\label{fig:ICARUS}
\end{figure}

\begin{figure}
\includegraphics[width=6.5in]{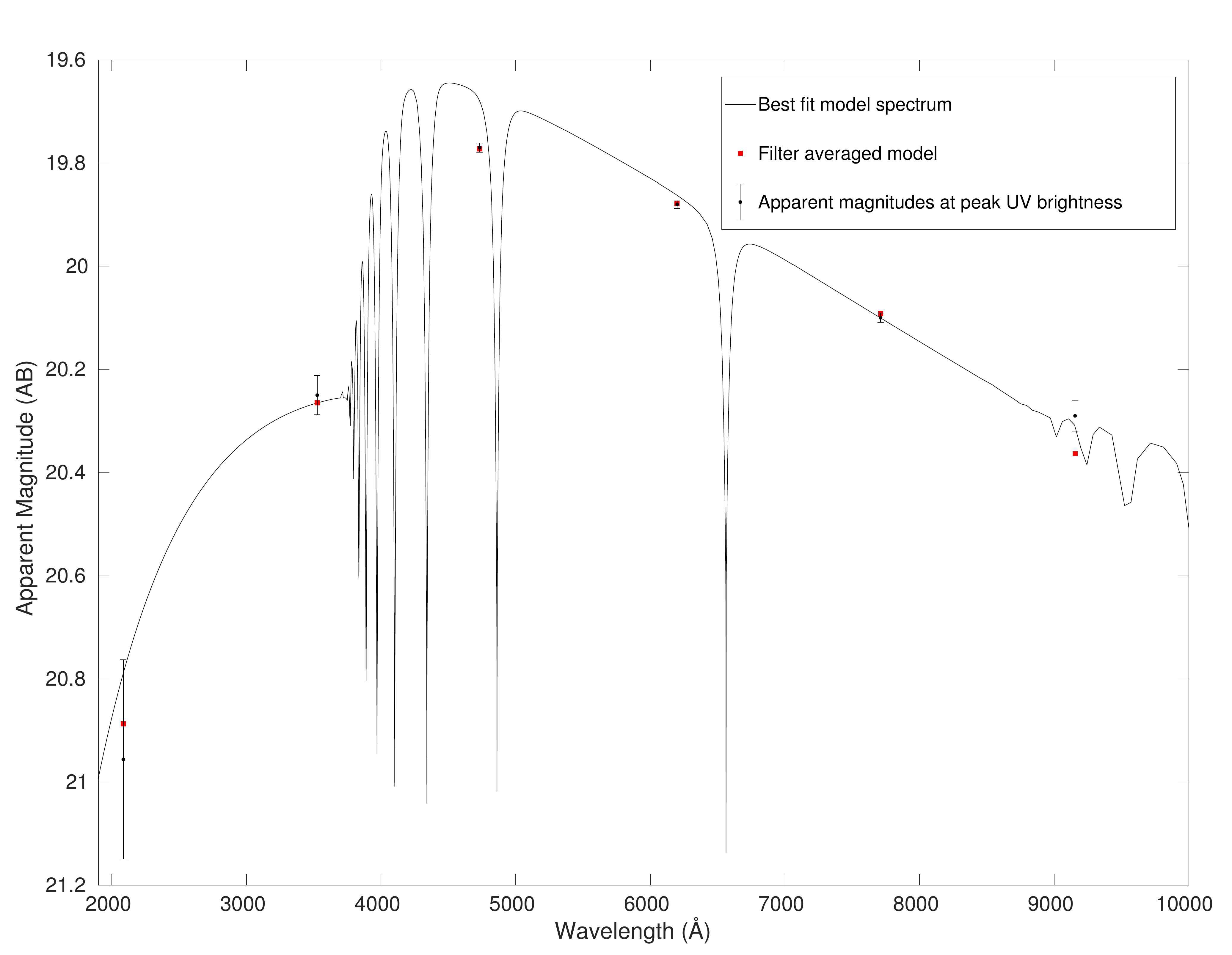}
\linespread{1.3}\selectfont{}
\renewcommand{\figurename}{Extended Data Figure}
\caption{\textbf{Model fit to peak of ZTF J1406+1222's spectral energy distribution} The spectral energy distribution at the orbital phase of the Swift UVOT maximum flux. The red squares illustrate the filter averaged model, which convolves the model spectrum with the respective HiPERCAM and Swift filters. The solid line is the best-fit white dwarf model spectrum. We used a grid of white dwarf model atmospheres because they naturally cover the appropriate surface gravities and temperatures needed to model the irradiated face of the companion, which could be either a brown dwarf or a white dwarf. The Swift detection strongly constrains its dayside temperature, giving a best-fit value of $T_{\mathrm{day}} = 10462\pm150\,\mathrm{K}$. All the error bars illustrated represent $1\sigma$ uncertainties in apparent magnitude.}

\label{fig:Peak_fit}
\end{figure}

\subsection{Spectroscopic analysis}

The spectroscopic data were obtained with LRIS using the 600/4000 grism with 2x2 binning on the blue arm and the 400/8500 grating with 2x1 binning on the red arm, on the nights of July 18 2020 UTC and Jan 10 2021 UTC. One the night of June 5 2021, we used the 600/3400 grism with 2x2 binning on the blue arm and the 600/7500 grating with 2x2 binning on the red arm. On this last night, we adjusted the position angle of the slit to $-$79.42\,degrees East of North in order to capture both sources and were also able to take advantage of a newly upgraded detector on the red arm. Thus, we base much of our spectroscopic analysis on this data set, which is superior to the other nights, though our radial velocity measurements made use of all three nights of data. We reduced all the spectroscopic data using the LPIPE pipeline\cite{Perley2019}.

\subsection{Black widow radial velocity analysis}

In order to measure radial velocity variations, we used a custom code which simultaneously fit six Lorentzian profiles plus a quadratic polynomial to the full spectrum, with the Lorentzians fitting the hydrogen lines at 4861.35, 4340.46, 4101.74, 3970.07, 3889.05 and 3835.40~\AA.

We attempted to fit both the hydrogen absorption and emission lines in order to measure radial velocity variations on the orbital period. However, the hydrogen absorption lines are broad and only visible over a short orbital phase interval, preventing us from extracting reliable radial velocity shifts from them. This is not surprising given that these absorption lines are blended with variable emission from the ablation wind, as well as underlying features of the sdK spectrum.

However, we detected appreciable radial velocity variations in the much narrower Balmer emission lines. As discussed in the main text, we found that these features are blueshifted after superior conjunction of the black widow, and we interpret this as a signature that they originate from a cometary tail spiraling around the irradiated companion, or potentially from the intrabinary shock in the system. Figure~\ref{fig:Trail} presents the best sine fit to the emission-line radial velocity curve, overplotted with the phase-folded trailed spectra.

\subsection{Cool subdwarf atmospheric analysis}
We identified the sdK nature of the hierarchical companion due to the presence of strong CaH absorption lines, a signature of these low metallicity late type stars\cite{Allard1995}. Typically, late type main-sequence stars are characterized in terms of line indices, which are measures of the strength of an absorption line relative to the continuum, the most important of which are the CaH and TiO indices\cite{Reid1995}. Analyzing the spectral indices of ZTF J1406+1222 presents a unique challenge because it is variable, with a non-negligible contribution from the inner binary at the wavelengths of these indices. In order to compensate for the variability, we compute the indices using a coadd of spectra taken within the faintest 20 percent of the inner binary orbit in order to minimize its contribution. This analysis yielded the spectral indices and derived parameters presented in Extended Data Table~\ref{tab:sdK}. We use the empirical relation\cite{Woolf2009} to determine the metallicity of the sdK star based on the indices found. One way to compute its temperature is using the CaH2 index\cite{Woolf2009}, where the effective temperature is given by $T_\mathrm{eff}=2696+1618\times CaH2$, which gives $T_\mathrm{eff}=4020\pm70 \rm \, K$. The uncertainty has been estimated by computing the scatter of the indices of the individual phase resolved spectra. Additionally, by fitting the BT-NextGen (GNS93) atmospheric models for low mass stars\cite{Allard2012} to the region around the TiO+CaH band (between $6250\AA$ and $7250\AA$) in the spectra taken during the faintest portion of the orbit, as illustrated in the inset of Extended Data Figure 5, we also obtain estimates of  $T_\mathrm{eff}=3800\pm30 \rm \, K$ and $\rm [Fe/H]=-1.16\pm0.084~dex$, though the uncertainties, derived from the covariance matrix of the fit, are sensitive to the slope of the continuum and dilution due to the black widow component, and thus are underestimated. Additionally, we use the Ca II $8542\,\rm \AA$ line originating from the sdK to estimate the systematic velocity of the system by fitting for velocities in individual spectra, and taking a weighted average of these measurements. We measure a systemic velocity of $\gamma=67\pm30\,\rm km\, s^{-1}$, where the uncertainty is dominated by the systematic uncertainty of the absolute wavelength calibration which we estimate by taking a simple barycentric corrected coadd of the individual spectra, and measuring the velocity of the Ca II $8542\,\rm \AA$ line in this coadd and comparing with the value yielded by the weighted average of fitting the individual epochs. We also verified that the scatter in the velocities estimated using the individual epochs does not track the 62-minute orbital period of the black widow component.

\begin{figure}
\includegraphics[width=6.5in]{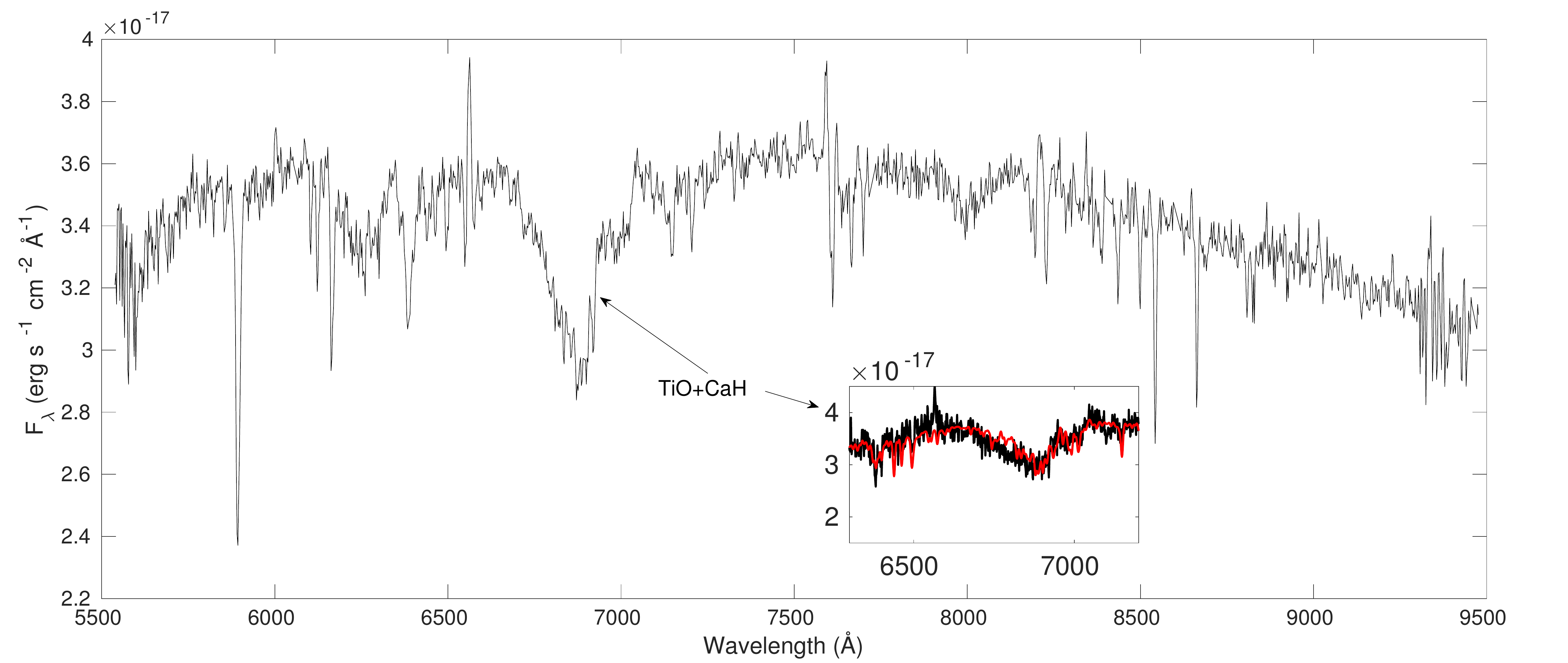}
\linespread{1.3}\selectfont{}
\renewcommand{\figurename}{Extended Data Figure}
\caption{\textbf{Red LRIS spectrum of ZTF J1406+1222} The red LRIS spectrum of ZTF J1406+1222 illustrating the significant contribution from the cool sdK component. The broad feature at $6700-7000\, \rm \AA$ is a combination of titanium oxide (TiO) absorption bands commonly seen in late type stars and strong calcium hydride (CaH) bands, which are more intense in low metallicity sdM/sdK stars. We classify the object as a low metallicity star by measuring the ratio of the TiO to CaH bands in a coadd of spectra at close to the minimum flux of the black widow component to avoid significant contamination of the continuum. Additionally, we also estimated the metallicity by performing a fit of atmospheric models to the region of the spectrum around the TiO+CaH band, as illustrated in the inset.}

\label{fig:Red_spectrum}
\end{figure}

\subsection{Distance Estimate}

We use three approaches to constrain the distance to the system: 1. using the parallax as reported in \emph{Gaia} eDR3\cite{Gaia2021}, 2. using the absolute magnitude of the sdK companion in the $K$ band, and 3. putting an upper limit based on the large proper motion of the system. We discuss all three approaches here.

As of Gaia eDR3, only the sdK component contains a full astrometric solution, and its parallax is measured to be $\varpi=2.1118\pm1.5882~ \rm mas$. Estimating a distance based on this is challenging, as the parallax measurement is hardly larger than its uncertainty. The best distance estimate based on this astrometric solution according to the Bailer-Jones et al. catalog\cite{Bailer-Jones2021} is $1290^{+2233}_{-724}\rm \, pc$. This value is poorly constrained, so we did not use it in our analysis, but instead used the properties of the sdK companion.

In order to estimate the distance to the system using the luminosity of the sdK, we interpolated over evolutionary model grids\cite{Baraffe1997} using the estimated metallicity $\rm [Fe/H]=-1.173~\rm dex$ and the observed colour $(V-K)=3.34\pm0.09$ as input parameters. Because our highest signal-to-noise ratio light curve was obtained in $g$, we empirically correct it to the $V$ band using the sample of objects presented in \cite{Jao2017}, which contains $g$ and $V$ magnitudes for a large sample of objects. We find that over this entire sample of objects, the $g-V$ correction factor is $0.738\pm0.072$. We use the minimum brightness of our HiPERCAM $g$ light curve of $21.289\pm0.0127 \rm \, mag_{AB}$ to estimate an apparent magnitude of the sdK of $V=20.55\pm0.09$. Technically, the black widow still contributes here, but if that were the case it would cause the sdK to appear warmer than expected and cause the distance to be slightly overestimated, which would drive lightcurve models even further in favor of a neutron star solution. The model grids also contain estimates of the effective temperature, which we found to be $T_\mathrm{eff}=3870 \pm 70~ \rm K$, consistent with the estimate made using the CaH2 index as well as with the spectroscopic model. The distance we calculated using this analysis and the observed UKIDSS $K$-band apparent magnitude of $K=17.182\pm0.054$ is $1140\pm200\,\rm pc$, with an approximate absolute $K$-band magnitude of $M_K=6.6\pm0.3$. This does not account for the uncertainty in the metallicity, as the grids used do not extend to metallicities above $\rm [Fe/H]=-1.0~dex$. Rather than extrapolate these grids, we made a more robust but broader estimate of the distance by investigating the measured absolute magnitudes of a nearby sample of cool subdwarf stars with colours similar to our observed colour of $(V-K)=3.34\pm0.09$. Based on the recent sample presented in \cite{Jao2017}, we find that the $K$-band absolute magnitude of the companion should fall between 6 and 8, in good agreement with the $M_K=6.6\pm0.3$ estimated using evolutionary models. The absolute magnitude range of $6<M_K<8$ corresponds to a distance interval of $680<d<1720 \rm \, pc$. Based on its absolute magnitude, spectral type, and metallicity, we estimate that the sdK has a mass of approximately $0.18\,M_{\rm \odot}$ to $0.3\,M_{\rm \odot}$\cite{Jao2016}.

Finally, in order to place a very conservative absolute upper bound on the distance using the proper motion, we assume that the cool companion should have a tangential velocity less than $1200 \rm ~ km \, s^{-1}$, as, if it exceeded this value, it would be the fastest hyper-velocity pulsar known\cite{Chatterjee2005}. This seems implausible, especially considering that it is a wide triple system. Given the $74.486\pm1.769~\rm mas\,yr^{-1}$ proper motion, this corresponds to a distance upper limit of approximately $3400\rm \, pc$. Additionally, an analysis of the object's kinematics revealed that it would be on a trajectory to escape the Galaxy if it were at a distance of $>2000\rm \, pc$, so the latter distance is a more realistic upper bound.

\begin{table}
\centering
 \caption{\textbf{Spectroscopic properties of the subdwarf K star}. The values presented here are shown in two columns, one considering indices of spectra taken within the faintest 30 percent of the light curve, centred around minimum brightness (orbital phases $\phi=0.8$ to $\phi=0.1$), and the other from a spectroscopic model fit to the coadd of the spectra taken around the faintest phase. The first four rows show the spectroscopic indices\cite{Reid1995} commonly used to characterize low metallicity, cool subdwarf stars \cite{Allard1995}. $\rm [Fe/H]$ is an estimate of the metallicity of the object based on the $\zeta_\mathrm{TiO/CaH}$ index computed according to the formalism described in \cite{Woolf2009}. The final row gives the metallicity classification of the object.}
\medskip
\begin{tabular}{lcc}
\hline
$\rm{Index/Parameter}$ &$\rm Value~(\phi=0)$ &$\rm Value~(Spectroscopic)$ \\
\hline
$ \rm{TiO5}$ & $0.953\pm0.044$ &  \\
$\rm{CaH2}$ & $0.820\pm0.041$ &   \\ 
$ \rm{CaH3}$ & $0.90\pm0.045$ &  \\
$ \rm{CaH2+CaH3}$ & $1.718\pm0.084$ &  \\
$ \zeta_\mathrm{TiO/CaH}$ & $0.45\pm0.50$ &  \\
$ \rm{[Fe/H]}$ & $-0.94\pm0.82$ & $-1.16\pm0.08$ \\
$ \rm{T_{\mathrm{eff}}}$ & $4020\pm70\,\rm K$ & $3800\pm30\,\rm K$ \\
$ \rm{Classification}$ & $\rm esdK$ & $\rm esdM$ \\

\hline
\end{tabular}
\label{tab:sdK}
\end{table}

\subsection{Establishing the hierarchical triple nature of the system}

Archival SDSS\cite{York2000} and Pan-STARRs1\cite{Chambers2016} images (Extended Data Figures 6 and 7) revealed the presence of what appears to be two distinct sources separated by a fraction of an arcsec, even though \emph{Gaia} DR2 only detected a single source at this position. With the release of \emph{Gaia} eDR3, a second source was identified by \emph{Gaia} $0.5555\pm0.0045$ arcsec away from the previously detected source. However, no further astrometric information is available for this source. The \emph{Gaia} source with a full astrometric solution exhibits a large proper motion of $74.486\pm1.769~\rm mas\,yr^{-1}$, which corresponds to the sdK star seen in the spectrum. By monitoring the position of the centroid of the point spread function (PSF) of the source in the HiPERCAM images, we were able to demonstrate that these two sources must be co-moving, and thus members of a hierarchical triple system. In Extended Data Figure 8, we illustrate that the position angle between the two sources has not changed significantly since the \emph{Gaia} measurement in 2016, whereas if one assumed that the variable black widow component was a distant background source with low proper motion and only the sdK exhibited this large proper motion, this position angle would have changed substantially in five years. Thus, we conclude that the sdK is a common proper motion companion to the 62-minute binary, making it the outer companion in a hierarchical triple system. 

\begin{figure}
  \centering
  \begin{minipage}[b]{0.45\textwidth}
    \includegraphics[width=\textwidth]{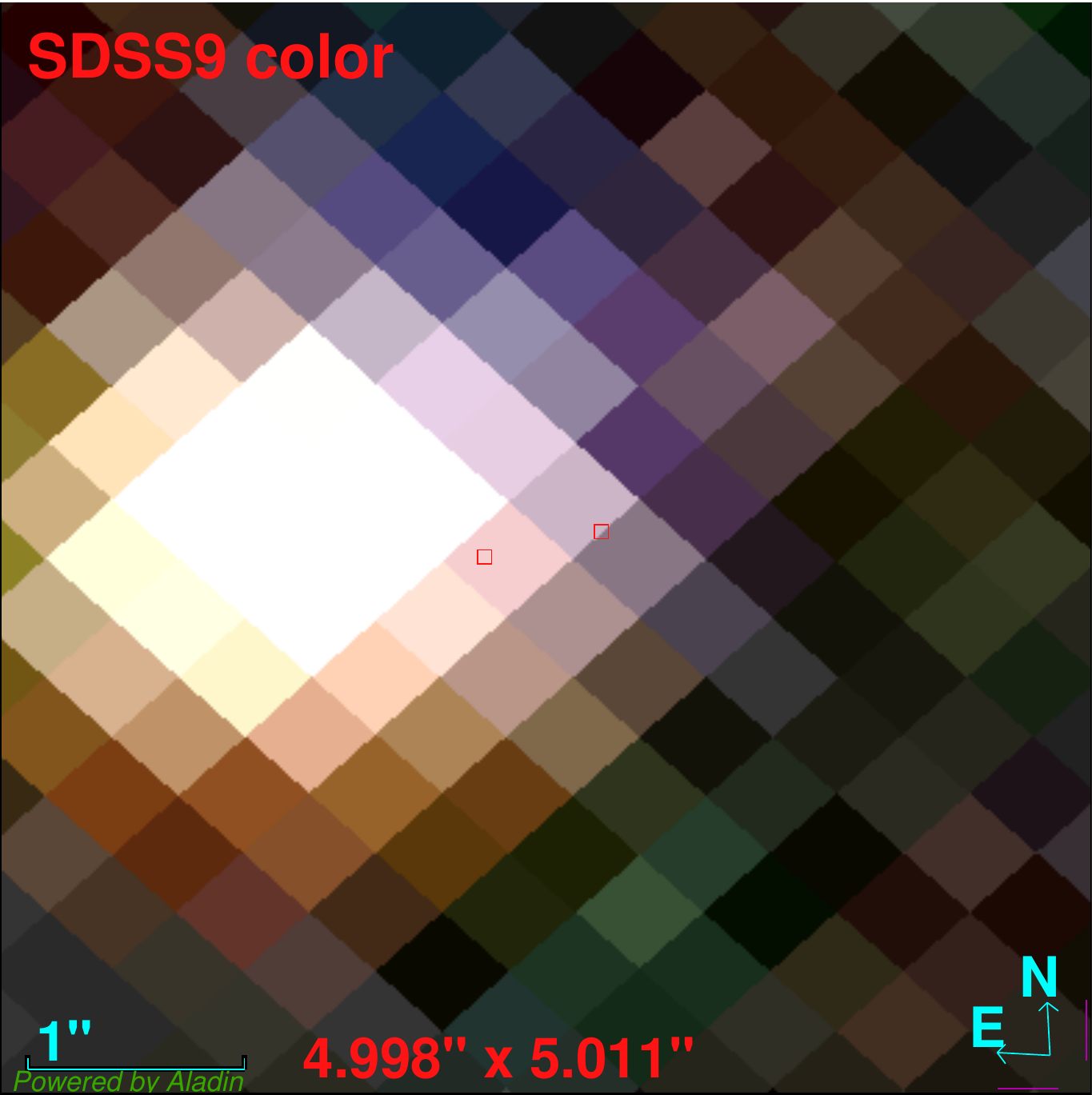}
    \linespread{1.3}\selectfont{}
    \renewcommand{\figurename}{Extended Data Figure}
    \setcounter{figure}{5}  
    \caption{\textbf{SDSS colour image of ZTF J1406+1222} An SDSS DR9 colour image cutout of the ZTF J1406+1222, revealing an asymmetric colour across the PSF with the sdK on the left. The two red boxes indicate the J2016.0 positions of the two \emph{Gaia} eDR3 sources. This image has been centred on the same coordinates as the Pan-STARRS1 cutout shown in Figure~\ref{fig:PS}.}
    \label{fig:SDSS}
  \end{minipage}
  \hfill
  \begin{minipage}[b]{0.45\textwidth}
    \includegraphics[width=\textwidth]{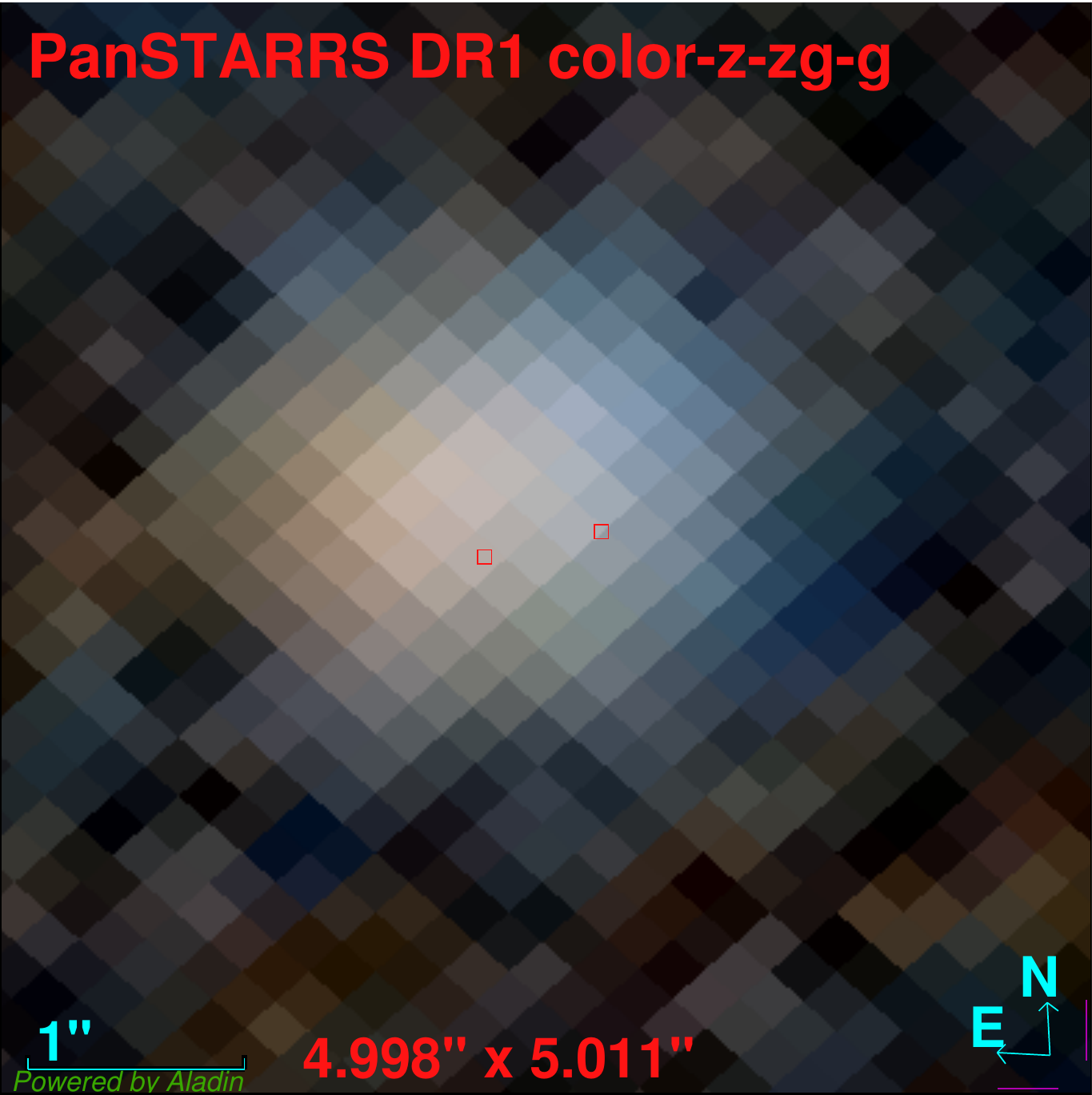}
    \linespread{1.3}\selectfont{}
    \renewcommand{\figurename}{Extended Data Figure}
    \caption{\textbf{Pan-STARRS1 colour image of ZTF J1406+1222} A Pan-STARRS1 colour image cutout at the same centroid as the SDSS image in Figure~\ref{fig:SDSS}. The Pan-STARRS1 PSF exhibits the same colour asymmetry as seen in the SDSS image, and the centroid is closer to the Gaia source position than the SDSS image due to the $74.5 \rm\, mas\,yr^{-1}$ proper motion of the system. }
    \label{fig:PS}
  \end{minipage}
\end{figure}

\begin{figure}
\includegraphics[width=6.5in]{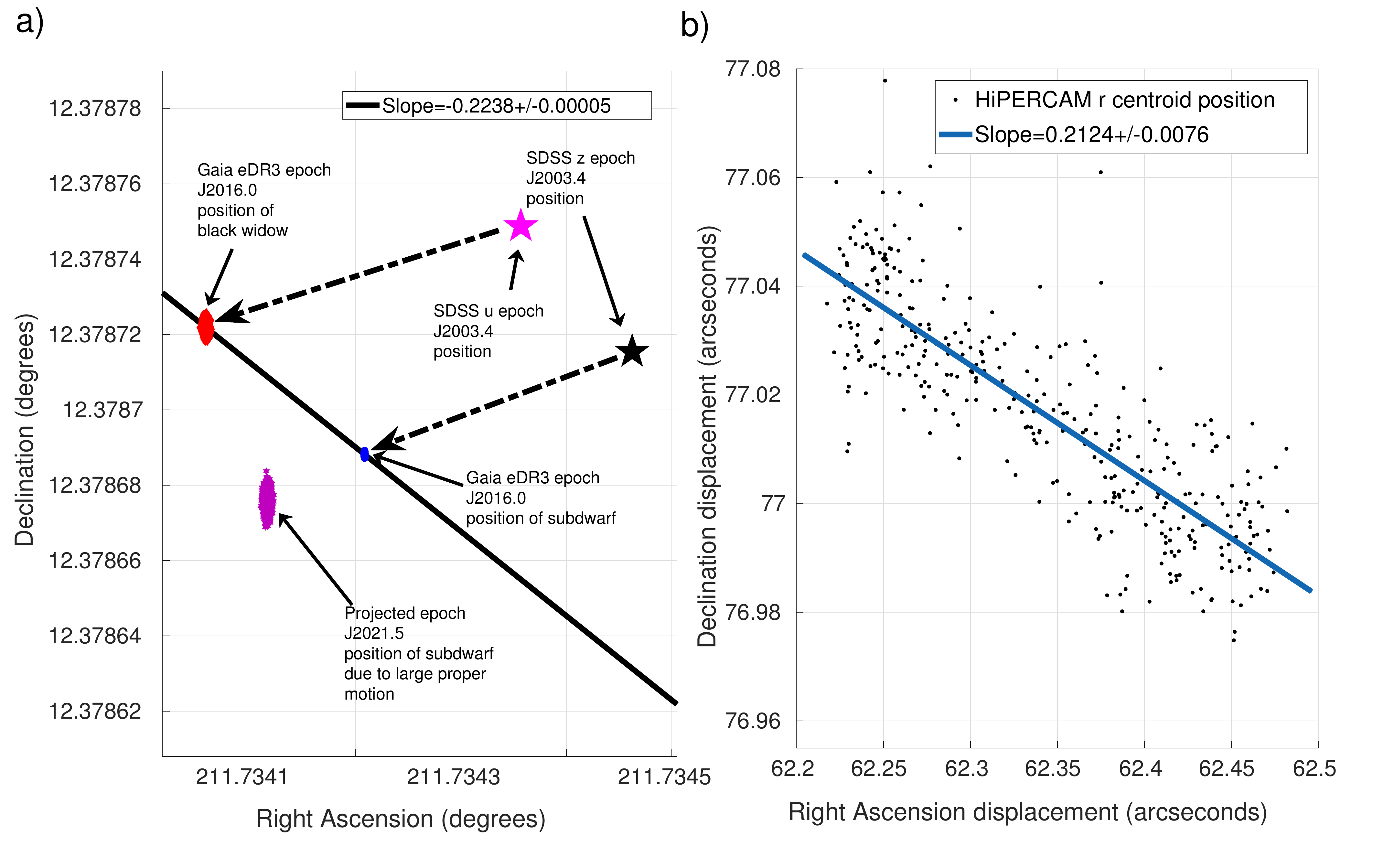}
\linespread{1.0}\selectfont{}
\renewcommand{\figurename}{Extended Data Figure}
\caption{\textbf{Astrometric characterization of ZTF J1406+1222} \textbf{a)} The measured \emph{Gaia} eDR3 J2016 position of the sdK is shown by the blue circles and the measured position of the variable black widow component is illustrated by the red diamonds. The magenta stars represent the projected J2021.5 position of the sdK, given the proper motion in its \emph{Gaia} astrometric solution. The solid black line shows the position angle between the two \emph{Gaia} source positions. The pink star indicates the position of the source in an SDSS u image obtained in J2003.4, and the black star indicates the source position in the SDSS z image at this epoch. Because the BW component dominates in u band, and the sdK in z band, these positions should approximately reflect the positions of the two components in J2003.4, and the dashed arrows indicate that these sources are co-moving, having both translated their positions significantly since that epoch (by about an arcsecond). \textbf{b)} The black points indicate the measured centroid of the point spread function of the variable source with respect to a reference star on the HiPERCAM $r$-band images. This centroid moves back and forth between the two sources of luminosity as the black widow component brightens and fades, and gives a precise estimate of the position angle between the two sources when the data were obtained, at epoch J2021.5. The solid blue line is a linear fit to the data used to derive a slope to measure this position angle, whose value is shown in the legend. The slope is consistent with the J2016 position angle and clearly inconsistent with only the sdK having moved since epoch J2016.0, demonstrating that the two sources must indeed be co-moving, and thus part of a hierarchical triple system.}
\label{fig:Position}
\end{figure}

We estimate the separation of the sdK and the BW by using the observed angular separation of $0.5555\pm0.0045$~arcsec between the \emph{Gaia} positions, and the estimated distance of $1140\,\rm pc$, yielding a separation of approximately $600$ astronomical units. Using the estimated sdK mass of $0.25\rm\,M_{\odot}$, and taking the inner black widow binary as having a mass of approximately $1.45\rm\,M_{\odot}$, we estimate that this would correspond to an orbital period of $11000$ years in a circular orbit (around $10500\pm500\rm\, years$ if one instead allows the mass of the inner BW to vary between $1.4\rm\,M_{\odot}$ and $2.0\rm\,M_{\odot}$ rather than fixing it to $1.45\rm\,M_{\odot}$.) It is quite possible that the orbit of the sdK is far from circular, which means we can only estimate a lower limit to the orbital period, by considering that the current separation represents the maximum separation of a highly elliptical orbit. 

\subsection{Swift XRT and NuSTAR analysis}

We obtained five separate Swift XRT observations of the object (observation IDs 00013598001, 00013598002, 00013598003, 00013598004, 00013598005) for a total of $12\, \rm ks$ exposure time. We used \texttt{XIMAGE} to coadd all the XRT exposures and compute a 3-$\sigma$ upper limit on the count rate within an 18 arcsec aperture using the \texttt{uplimit} routine of \texttt{XIMAGE}. The aperture only contained 1 count, and the 3-$\sigma$ upper limit on the count rate was $6.648\times10^{-4}\, \rm count~s^{-1}$. We used \texttt{WebPIMMS} to convert this into a 3-$\sigma$ upper limit on the unabsorbed flux in the $0.2-10~\rm keV$ band, which is $2.6\times 10^{-14}\,\rm erg\,cm^{-2}\, s^{-1}$, assuming a column density of $n_{\mathrm H}=2\times 10^{20}$ along the line of sight. Taking our estimated distance of $1140\,\rm pc$, this translates into a 3-$\sigma$ upper limit on the source luminosity of $4.04\times10^{30}\,\rm erg\,s^{-1}$ at $0.2-10\,\rm keV$. We assumed a power law spectrum with a photon index of $2$ when estimating the source flux.

We also obtained a NuSTAR observation (ID 90601325002) with $28\,\rm ks$ on source time with module A and $25\,\rm ks$ with module B. The source was not detected, with 191 counts in a $50$ arcsec aperture in module A, with an average background of 177 counts determined using an annulus centred on the source with an inner and outer radius of $100$ and $200$ arcsec, respectively. The $50$ arcsec aperture in module B produced 192 counts and the background amounts to an average of $205$ in an equivalent area. Using \texttt{XIMAGE}, we summed the two exposures and supplied the summed background rate to the \texttt{uplimit} routine, which yielded a 3-$\sigma$ upper limit of $0.0013899\,\rm count\,s^{-1}$. Using \texttt{WebPIMMS}, we estimate this corresponds to an unabsorbed flux of $1.178 \times 10^{-13}\,\rm erg\,cm^{-2}\, s^{-1}$, translating to a luminosity upper limit of $1.83 \times 10^{31}\,\rm erg\,s^{-1}$ at $0.2-10\,keV$ assuming a distance of $1140\,\rm pc$, and a column density of $n_{\mathrm H}=2\times 10^{20}$ along the line of sight. Like with the Swift observation, we assumed a power law spectrum with a photon index of $2$.

There is a relation between the pulsar spin down luminosity, $\dot{E}$, and the heating luminosity inferred from the irradiated companion, $L_\mathrm{H}$, and the X-ray luminosity in the system\cite{Lee2018}. In Extended Data Figure 9, we illustrate the deeper Swift upper limit and our estimates on the $\dot{E}$ based on light curve modelling and assuming a pulsar heating efficiency factor $0.1<\eta<1$. The black diamonds indicate the measured X-ray luminosity of other black widow systems, and the red triangles are upper limits for other black widows which have not yet been detected in X-rays. Our Swift observation would have failed to recover several known systems at the estimated distance of $1140\,\rm pc$ and similar $\dot{E}$, and thus future deeper X-ray observations will be informative in further constraining ZTF J1406+1222's X-ray luminosity. Unlike other black widow systems, we have not directly detected the pulsar, and thus have no direct measurement of $\dot{E}$, but rather only an estimate of $L_\mathrm{H}$. Typical $L_\mathrm{H}$ values in spider binaries are on the order of $\eta=0.15$, indicating that only a fraction of the spin down luminosity of the neutron star contributes to heating the companion. However, for illustrative purposes, we choose an upper limit of $\eta=1$ because ZTF J1406+1222, with its extremely short orbital period and cometary tail most resembles PSR J1311-3430, an object consistent with $\eta>1$. Light curve models are unable to fully capture the complicated heating physics in such systems, for example, an intrabinary shock which wraps around the companion and subtends a larger cross section of the neutron star spin down flux than possible with only the companion.
\begin{figure}
\includegraphics[width=6.5in]{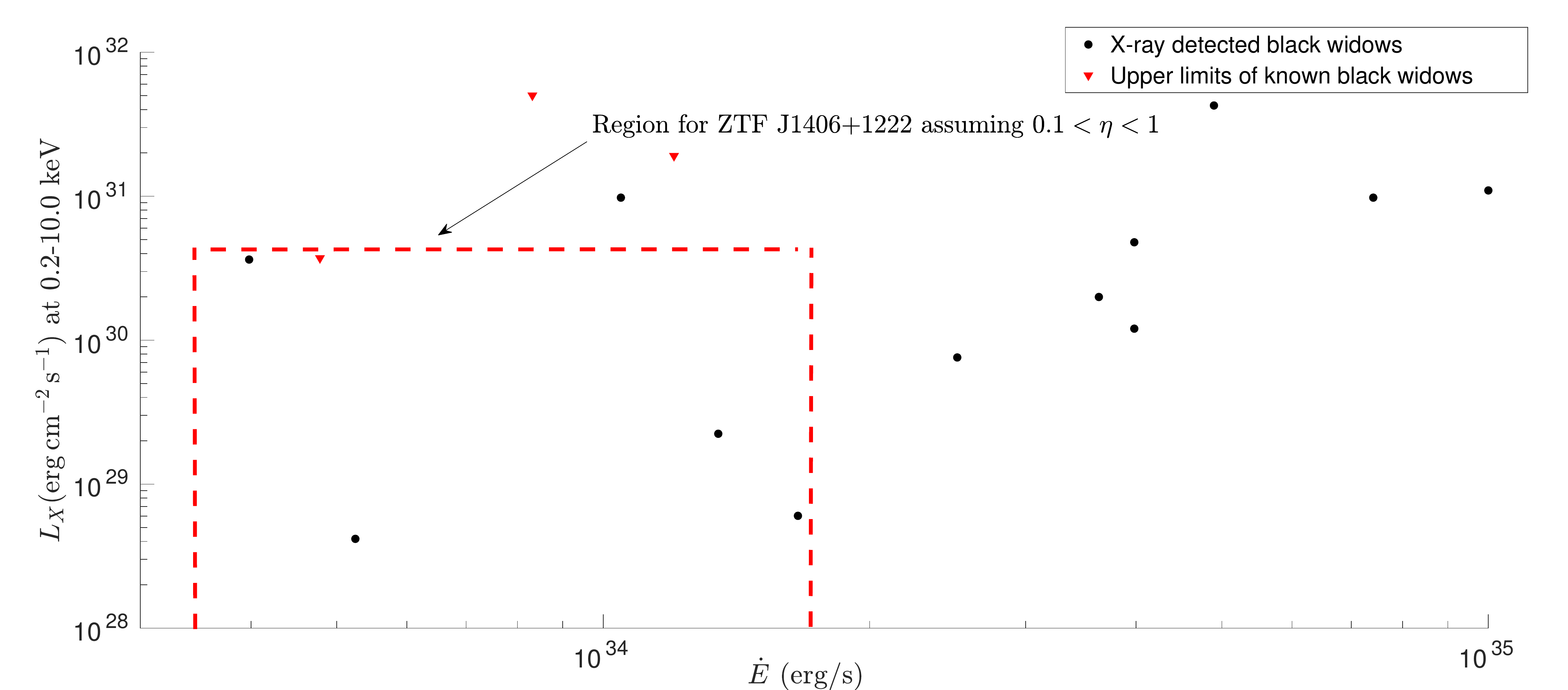}
\linespread{1.0}\selectfont{}
\renewcommand{\figurename}{Extended Data Figure}
\caption{\textbf{X-ray luminosity constraint of ZTF J1406+1222} The $0.2-10\,\rm keV$ X-ray luminosity, $L_\mathrm{X}$ vs $\dot{E}$ for known black widow systems\cite{Lee2018,Kong2014} (detections shown as black circles and upper limits as red triangles). The red dashed line indicates the region which we have constrained ZTF J1406+1222 to occupy based on the X-ray luminosity upper limit derived from our Swift observation and the estimate of $\dot{E}$ based on the peak temperature of the irradiated face of the neutron star's companion. Several known millisecond pulsars with a similar $\dot{E}$ are currently below our X-ray flux upper limit, and thus deeper observations may yield an X-ray detection of ZTF J1406+1222.}
\label{fig:xray}
\end{figure}

\subsection{Swift UVOT analysis}

We constructed a UVW2-band light curve of the system from the Swift observations 00013598003, 00013598004 and 00013598005. We used the \texttt{uvotlc} tool to bin event mode data into $10\,\rm s$ snapshots and constructed a binned light curve by computing a weighted sum of the measured fluxes in these snapshots according to orbital phase. These measurements are presented in Extended Data Figure 2, where we have omitted $<1$-$\sigma$ detections (many of which were consistent with negative fluxes and would not have translated to a magnitude scale). One snapshot, 00013598005, was centred around the time of minimum light with an exposure duration of $1736\,\rm s$, covering the full fainter half of the orbit. The source was not detected above background in this deep exposure, and we used XIMAGE and a $5$ arcsec aperture to compute a 3-$\sigma$ upper limit on the apparent magnitude of the object during the fainter half of the orbit, which we found to be $>22.9\,m_\mathrm{AB}$ in the UVW2 band. 

\subsection{Fermi Analysis}

The source is not detected by the Fermi Gamma-ray Telescope. We computed an upper limit to the gamma-ray luminosity using a prescription similar to that used for Swift~ J1644+5734\cite{Burrows2011}. We considered a 15-degree region around the source and the $100-10000\,\rm MeV$ energy range, and perform a binned likelihood analysis of Fermi LAT data between 2008-08-04 and 2021-08-27. We find a 3-$\sigma$ upper limit on the photon rate of $9.64\times10^{-10}\rm \, photon\,cm^{-2}\,s^{-2}$. We assumed a power law with photon index 2, which is typical of spider binaries at gamma-ray energies\cite{Hui2019}. This corresponds to a $3\sigma$ uppper limit on the gamma-ray luminosity of $L_{\gamma}<1.45\times10^{32}\,\rm erg\,s^{-1}$ for an assumed distance of $1140\,\rm pc$; this means the source is fainter than most known black widow binaries in the Fermi 4FGL catalog exhibit\cite{Hui2019}. The system is at a similar distance to PSR J0636+5128, a 95.8-minute orbital period black widow which is also not in the Fermi LAT 8 yr catalog\cite{Draghis2018}. As discussed in the work on PSR J0636+5128\cite{Draghis2018}, this may be due to gamma-ray luminosity being preferentially beamed towards the spin equator in the system, our our viewing the system at high inclination. Our heating models of the lightcurve are able achieve good fits with physical solutions for inclinations as low as 35 degrees, and thus it is possible that for if the spin of the pulsar is aligned with the binary orbit, we intercept very little gamma-ray luminosity. Because the ZTF J1406+1222 was selected optically, its selection was not biased towards gamma-ray bright sources like those selected using Fermi.

\subsection{Galactic kinematic analysis}

The high proper motion and low metallicity sdK companion of ZTF J1406+1222 are suggestive that the system is an object in the Galactic halo. To confirm this, we used the \texttt{galpy}\cite{Bovy2015} package to compute its trajectory around the Milky Way over 6 Gyr, using the McMillan2017 potential\cite{McMillan2017}, and it's Gaia astrometric solution and the radial velocity of the sdK measured from the LRIS spectra. The resulting trajectories, shown in Extended Data Figures 10 and 11, are clearly consistent with a halo trajectory in which the object reaches scale heights of $>10\,\rm kpc$ out of the Galactic disk, as seen in Figure~\ref{fig:Orbit1}, and $>10\,\rm kpc$ radially from the Galactic centre, as seen in Extended Data Figure 11. This analysis revealed that the object would escape the Galaxy when placed at a distance of $>2000\,\rm pc$. Significantly, the solution corresponding to ZTF J1406+1222 being 940~pc away from Earth passes within just 50 parsecs of the Galactic center, as illustrated in Extended Data Figure 11. This suggests that the system may have originated from a disrupted globular cluster.

\begin{figure}
\includegraphics[width=6.5in]{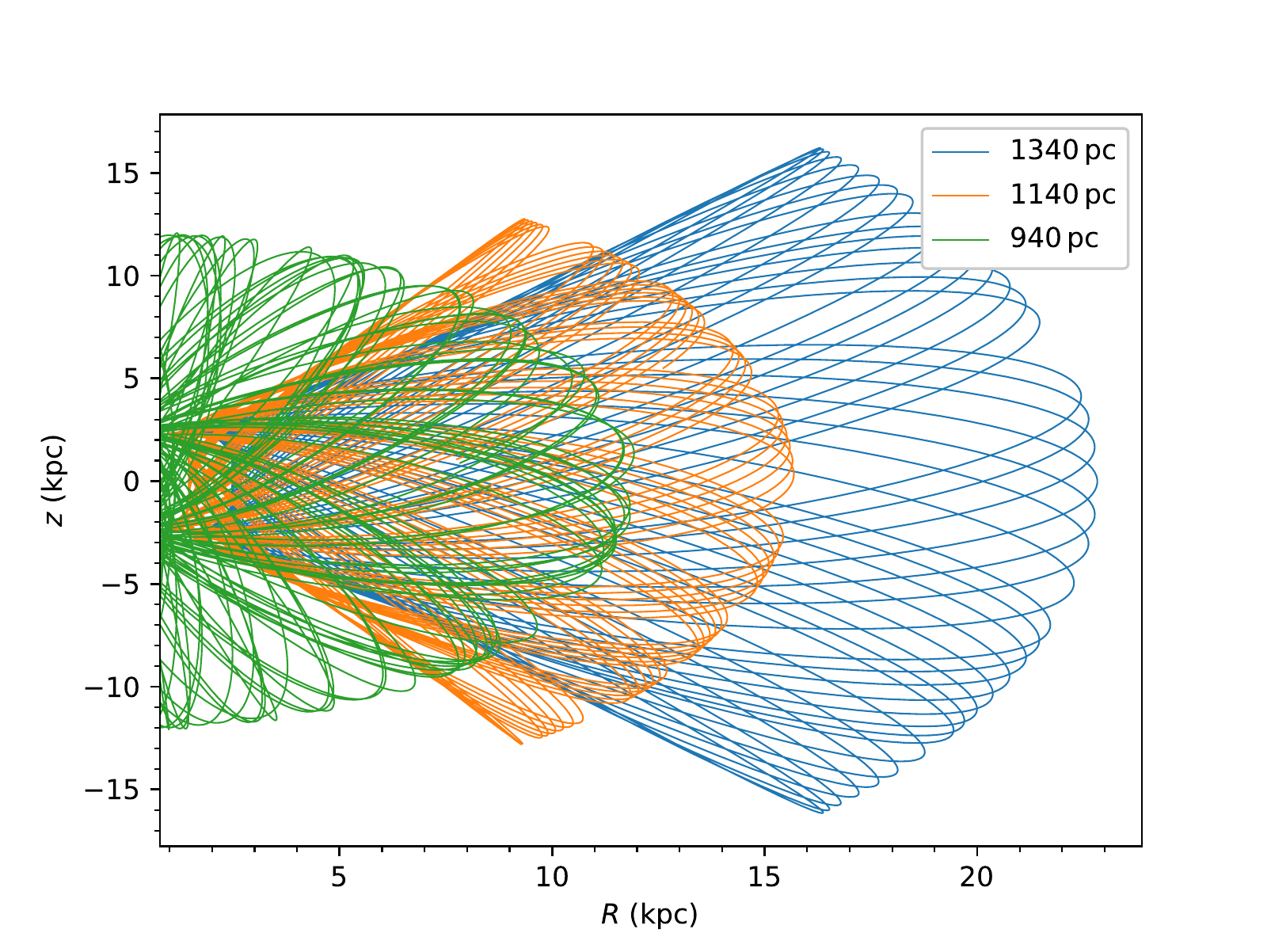}
\linespread{1.3}\selectfont{}
\renewcommand{\figurename}{Extended Data Figure}
\caption{\textbf{Radial vs scale height orbital solution of ZTF J1406+1222 around the Galaxy} A plot illustrating the Milky Way orbit of ZTF J1406+1222 over the course of 10 Gyr. The colours (green, orange and blue) indicate three different distances corresponding to our best distance estimate of $1140\pm200\,\rm pc$ and the 1-$\sigma$ upper and lower bounds of this distance estimate. In all cases, the object reaches a scale height ($z$) of more than $10 \, \rm kpc$ above the Galactic disk and travels a great distance ($R$) away from the Galactic centre in the radial direction, clearly indicating that it is a halo object. Notably, the green line, illustrating the orbital solution if ZTF J1406+1222 is at  $\approx 940\pm200\,\rm pc$ passes within 50 parsecs of the Galactic center.}

\label{fig:Orbit1}
\end{figure}

\begin{figure}
\includegraphics[width=6.5in]{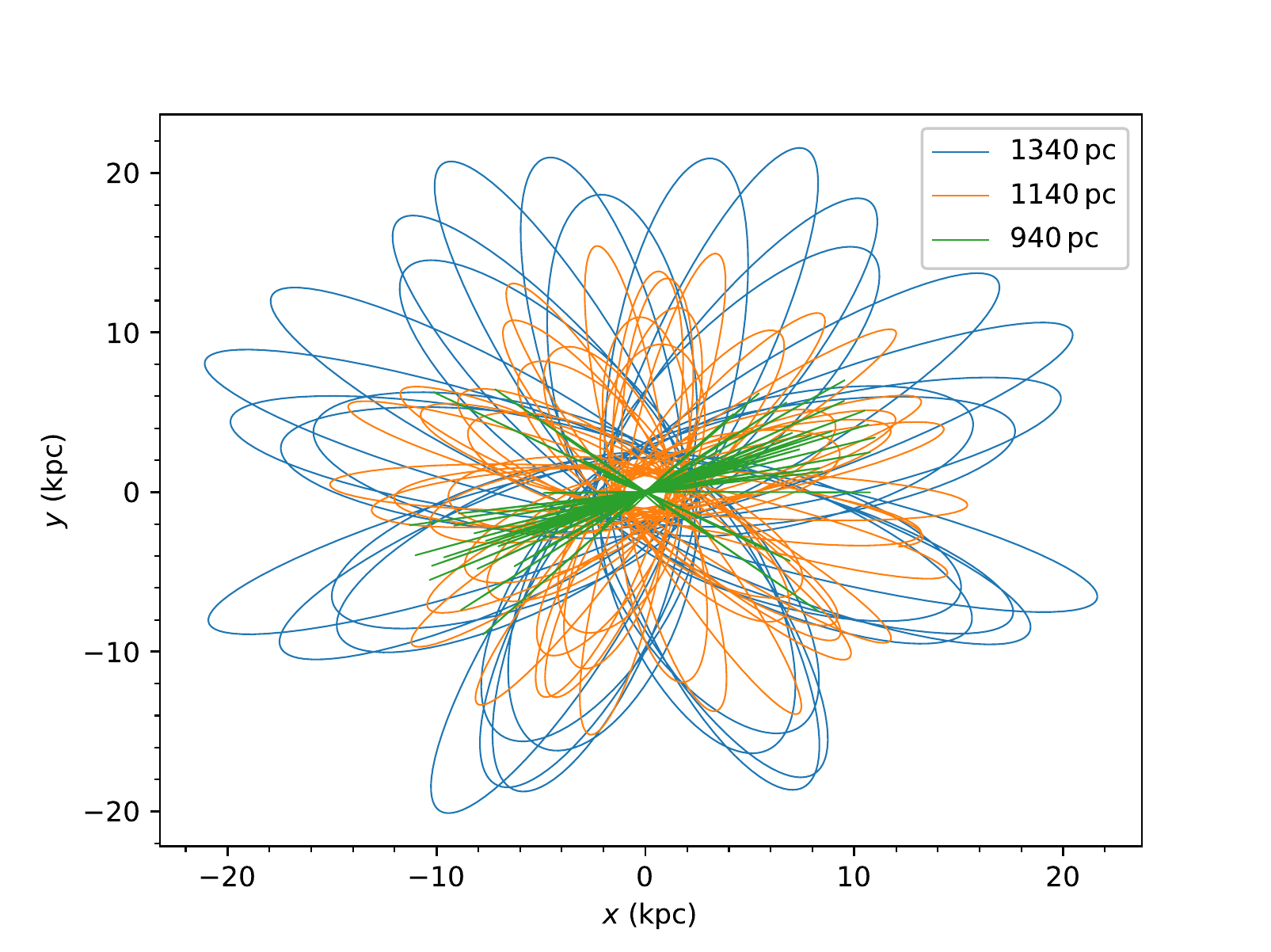}
\linespread{1.3}\selectfont{}
\renewcommand{\figurename}{Extended Data Figure}
\caption{\textbf{Cross section of ZTF J1406+1222's orbital solution in the Galaxy} Cross section of the Milky Way orbit of ZTF J1406+1222 over the course of 10 Gyr. The colours (green, orange and blue) indicate three different distances corresponding to our best distance estimate of $1140\pm200\,\rm pc$ and the 1-$\sigma$ upper and lower bounds of this distance estimate. }

\label{fig:Orbit2}
\end{figure}

\subsection{Searches for Radio Pulsations}

Black widow pulsars are capable of producing detectable radio pulsations (e.g. see~\cite{Fruchter+1988, Manchester+1991, Deich+1993, Ransom+2005, Ray+2013}), provided the radio emission is beamed toward Earth and it is not scattered or absorbed. These effects can be mitigated by performing observations at higher radio frequencies. To search for radio pulsations from ZTF~J1406+1222, we performed three radio observations with the NASA Deep Space Network~(DSN; see~\cite{Pearlman+2019a}) radio telescopes, which were carried out with the DSS-13 (34\,m diameter) and DSS-14 (70\,m diameter) antennae. The DSS-13 observation (epoch~1) was carried out for 56 minutes starting at 2020~July~02 21:57:00~UTC, at a center frequency of 2.26\,GHz ($S$-band) with a recording bandwidth of 110.625\,MHz. Two additional observations were performed with DSS-14 during separate epochs, with one starting at (epoch~2) 2020~July~13 22:02:24~UTC for a duration of 183~minutes at a center frequency of 1.54\,GHz ($L$-band) with a recording bandwidth of 320.625\,MHz, and another starting at (epoch~3) 2021~July~31 05:35:00~UTC for a duration of 75~minutes at a center frequency of 2.24\,GHz ($S$-band) with a recording bandwidth of 115.625\,MHz. During each observation, power spectral measurements were recorded across the band at high time resolution in a digital polyphase filterbank with a sampling time of 102.4\,$\mu$s and a frequency resolution of 0.625\,MHz.

The data processing procedures are similar to those described in earlier studies of pulsars and magnetars with the DSN~(e.g. see~\cite{Majid+2017, Pearlman+2018a, Pearlman+2019a, Pearlman+2020b}). We reduced the radio data by first identifying and masking data corrupted by radio frequency interference~(RFI) using the \texttt{rfifind} tool available in the \texttt{PRESTO} pulsar search package\cite{Ransom2001}. The data were then bandpass-corrected, and low frequency fluctuations in the baseline were removed by subtracting the moving average from each data point in each frequency channel, which was calculated using 100~ms of data around each time sample. Next, the sample times were corrected to the solar system barycenter~(SSB) using the \texttt{prepdata} tool available in \texttt{PRESTO} and JPL's DE405 ephemeris.

The maximum predicted DM contribution along the line-of-sight is 23.4/21.9\,pc\,cm$^{\text{--3}}$ according to the NE2001/YMW16 electron density models\cite{Cordes+2002, Yao+2017}, and there is likely an additional, unknown contribution to the DM from the dense, ionized plasma wind produced by evaporation of the companion. We chose to incoherently dedisperse the $L$-band and $S$-band data using dispersion measure~(DM) trials between 0 and 5000\,pc\,cm$^{\text{--3}}$. For each observation, the DM trial spacing was chosen to minimize the total dispersive smearing (e.g. due to dispersive smearing within each frequency channel, dispersive smearing across all frequency channels from the DM step-size, and the sampling time). When determining the dedispersion scheme, we ignored the effects of scattering, which would have resulted in larger DM step-sizes.

We searched for bright, astrophysical single pulses using a Fourier domain matched-filtering algorithm, where each dedispersed time series was convolved with boxcar functions with logarithmically spaced widths between 102.4\,$\mu$s and 30.72\,ms. Candidates identified from each DM trial with a signal-to-noise ratio~(S/N)\,$\geq$\,6 were saved and classified using the \texttt{FETCH} (Fast Extragalactic Transient Candidate Hunter) software package\cite{Agarwal+2020}, which uses a deep-learning convolutional neural network to identify astrophysical radio pulses. No astrophysical single pulses were detected during any of our radio observations. We place the following 6$\sigma$ upper limits on the peak flux density of single pulses from ZTF~J1406+1222, assuming a fiducial width of 1\,ms: (epoch~1) $<$\,1.88\,Jy at $S$-band, (epoch~2) $<$\,0.23\,Jy at $L$-band, and (epoch~3) $<$\,0.41\,Jy at $S$-band.

We also carried out a search for radio pulsations using \texttt{accelsearch}, a Fourier Domain Acceleration Search~(FDAS) pipeline available in \texttt{PRESTO}, which employs a matched-filtering algorithm to correct for Doppler smearing. The search was carried out by summing 16 harmonics and using the \texttt{-zmax 1200} option, which defines the maximum number of Fourier bins that the highest harmonic can linearly drift in the power spectrum (e.g. due to orbital motion). We carried out pulsation searches at $L$-band and $S$-band by independently searching overlapping data segments spanning 10\% of the orbital period (6.2\,min). Searching shorter data segments, spanning 10\% of the orbit, aids in mitigating the sensitivity loss due to orbital motion since the pulsar spin frequency experiences an approximately linear drift in the regime where $T_{\text{obs}}$\,$\lesssim$\,$P_{\text{orb}}$/10~\cite{Ransom+2003}. We also carried out searches for pulsations with periods between 1\,ms and 100\,s in each dedispersed time series segment using a GPU-accelerated Fast Folding Algorithm~(FFA). We folded the dedispersed data modulo each of the period candidates identified by the two algorithms, but we found no statistically significant signals with S/N\,$\geq$\,6 in any individual data segment. We place the following 6$\sigma$ upper limits on the radio flux density of ZTF~J1406+1222, assuming a duty cycle of 10\% and an integration time of 6.2\,min: (epoch~1) $<$\,0.7\,mJy at $S$-band, (epoch~2) $<$\,0.1\,mJy at $L$-band, and (epoch~3) $<$\,0.2\,mJy at $S$-band. These sensitivity limits are comparable to the typical luminosities of known pulsars at 1~kpc in the ATNF pulsar catalog\cite{Manchester2005}, and many pulsars in the catalog at this distance are below this luminosity threshold.

There are several possible explanations for the lack of radio pulsations observed from this system. It is possible that the black widow pulsar is either radio-quiet, producing radio pulsations that are below the above-mentioned detection thresholds, or its radio emission is not beamed toward Earth. Alternatively, if the pulsar is an aligned rotator, then it is unsurprising that pulsations were not be detected. At a distance of 1.14\,kpc, the predicted diffractive interstellar scintillation~(DISS) bandwidth is $\sim$8\,MHz and the predicted DISS timescale is $\sim$13~minutes at 1\,GHz (assuming a transverse velocity of 100\,km\,s$^{\text{--1}}$), according to NE2001\cite{Cordes+2002}. At $S$-band, the DISS bandwidth and DISS timescales are $\sim$177\,MHz and $\sim$30~minutes, respectively, assuming a scaling of $\Delta\nu_{\text{DISS}}$\,$\propto$\,$\nu^{\text{4}}$ and $t_{\text{DISS}}$\,$\propto$\,$\nu^{\text{1.2}}$. Since the scintillation bandwidth at $S$-band is larger than our observing bandwidth, it is possible that scintillation-induced modulation could have reduced the apparent flux density of the radio pulses to a level below our detection threshold. Since ZTF~J1406+1222 is a compact system, with an estimated semi-major axis of 0.59\,$R_{\odot}$ for a 1.4\,$M_{\odot}$ neutron star in a circular orbit with a 0.05\,$M_{\odot}$ companion, extreme scattering and eclipses due to the presence of gas flowing out from the irradiated companion is also likely a contributing factor to the detectability of the pulsar's radio emission.

\subsection{Formation Questions}

The wide tertiary companion and short inner binary orbital period of ZTF J1406+1222 are highly unusual amongst neutron star systems, prompting the question of whether these two peculiarities are related, and challenging formation models. The orbital period of 62 minutes is the shortest known for a black widow system, the previous record holder being PSR J1653.6-0159 with an orbital period of 75~minutes\cite{Romani2014}. Similar to cataclysmic variables, formation models for black widow and redback systems (e.g., \cite{Chen2013,Benvenuto2015,Ginzburg2020}) predict that hydrogen rich donors reach a minimum orbital period at $P_{orb} \approx 80 \, {\rm min}$, slightly longer than that of ZTF J1406+1222. Shorter minimum orbital periods are possible if the donor star is somewhat evolved\cite{Podsiadlowski2003,Benvenuto2014} (i.e., it has a helium-rich composition). Hence, the ablated component of ZTF J1406+1222 may have formed from a helium-rich star that began mass transfer near the end of its progenitor's main sequence evolution.

There is no obvious reason to believe the widely separated tertiary allowed the system to reach unusually short orbital periods. The dynamics of the inner binary would naively be decoupled from the influence of the outer tertiary once general relativistic precession occurs on a timescale shorter than the precession induced by the outer body, which will happen at orbital periods far greater than one hour. Similarly, any formation models involving tidal effects or mass transfer from the third body would probably not occur due to its very wide separation. The same applies to formation models invoking a three-body common envelope event\cite{Sabach2015}.

In fact, the wide tertiary companion in ZTF J1406+1222 poses serious problems for most formation models of the system. It is difficult to understand how the widely separated companion, with an orbital velocity of $\sim \! 1 \, {\rm km~s^{-1}}$, was not unbound due to a kick imparted to the inner binary upon the formation of the neutron star. In formation models of the pulsar triple system PSR J0337+1715\cite{Tauris2014}, the outermost orbit has a period of $P \! \approx \! 17 \, {\rm d}$ (orbital velocity of $\sim \! 100$ km/s) at the time of the supernova, so the system would have more easily stayed bound. Appealing to extremely small supernova kicks\cite{Igoshev2019} is not sufficient for ZTF J1406+1222 to have remained bound, as explained below.

In standard binary formation channels of spider systems and low-mass X-ray binaries, the binary evolution begins with a massive ($M_1 \sim 10-20 \, \rm M_\odot$) star at moderate orbital separation ($a \sim 1 \, {\rm AU}$) with a low-mass ($M_2 \sim 1 \, \rm M_\odot$) companion star. A common envelope phase occurs after the primary expands into a red supergiant, after which the binary is composed of the helium core ($M_1 \sim 3-6~\rm M_\odot$) of the primary in a $\sim \! 1 \, {\rm day}$ orbital period with the low-mass companion. The helium star undergoes a core-collapse explosion to form a neutron star, ejecting its remaining helium/carbon/oxygen envelope, and possibly being kicked in the process.

In the absence of significant envelope stripping via case-BB mass transfer after core He-burning\cite{Tauris2015} (as expected for higher mass helium stars), a few solar masses of material is expected to be ejected during the explosion. This is comparable to the remaining mass of the system, and thus the inner binary could become unbound due to the supernova mass loss, even in the absence of a kick. The instantaneous mass loss will also kick the centre of mass of the inner binary by a velocity $\Delta v = v_{\rm He} (M_{\rm ej}/M_{\rm final})$, where $v_{\rm He} = \sqrt{G(M_1+M_2)/a}[M_2/(M_1+M_2)]$ is the helium star orbital velocity before the explosion, $M_{\rm ej}$ is the ejecta mass and $M_{\rm final} = M_1 + M_2 - M_{\rm ej}$ is the post-explosion mass of the system. For a typical pre-explosion configuration of $M_1 \sim 4 \, \rm M_\odot$, $M_2 \sim 1 \, \rm M_\odot$, $M_{\rm ej} \sim 2.5 \, \rm M_\odot$ and $a \sim 7 \, \rm R_\odot$, we thus expect a kick of $\Delta v \sim v_{\rm He} \sim 70 \, {\rm km~s^{-1}}$. This is far larger than the orbital velocity of the low-mass tertiary object in ZTF J1406+1222, meaning that the third body should become unbound for any choice of parameters near our fiducial values.

There are a few ways in which the net kick could be lower than these estimates. If the companion mass $M_2$ at the time of core-collapse is much smaller, e.g. $M_2 \sim 0.1 \, \rm M_\odot$, this would lower the kick velocity by a factor of $\sim \!10$ to several km~s$^{-1}$, but this would likely still unbind the tertiary. If the ejecta mass was much smaller, e.g. $M_{\rm ej} \sim 0.25 \, \rm M_\odot$ as would be expected for a low-mass helium star that lost most of its envelope via case-BB mass transfer\cite{Tauris2015}, this would lower the kick by a factor of $\sim \! 10$, but once again the tertiary would likely become unbound. A combination of these two mechanisms could potentially operate, but one would expect the companion to accrete some of the transferred mass such that its pre-explosion mass increases well above $0.1 \, \rm M_\odot$. The mass loss kick could be diminished if the pre-explosion orbital period is longer than one day, but it can only be a few times larger in order to ensure that subsequent magnetic braking can shrink the system to the observed short period, which is not enough to greatly decrease the kick velocity. Finally, it is possible that a supernova kick to the neutron star could cancel out the mass loss kick so that the overall kick to the inner binary is very small. But given the typical kick scale of a few hundred kilometers per second, great fine tuning would be required to reduce the binary motion to less than a few km~s$^{-1}$. 

Since none of the solutions above are particularly compelling, other formation mechanisms should be considered. One possibility is that the inner binary of ZTF J1406+1222 had a much larger orbital separation than $a \sim 7 \, \rm R_\odot$, e.g., if it never went through a common envelope phase. This would entail a much smaller kick due to mass loss upon core-collapse, perhaps allowing the tertiary to remain bound. Given the right orbital inclination of the tertiary, the inner binary could be hardened by high-eccentricity migration\cite{Naoz2016} (e.g. Kozai oscillations). Upon tidal orbital circularization to short periods, magnetic braking and gravitational wave emission would then bring the system to the short orbital period observed. However, this mechanism would still require an exceptionally small supernova kick (less than a few km~s$^{-1}$) to the neutron star in order for the tertiary to remain bound, which may be unlikely. 

Another possibility is that this system was dynamically assembled and ejected from a globular cluster. In order to dynamically eject a bound triple system, a four-body interaction would be required. For instance, a binary-binary interaction could have occurred, ejecting one star and forming a bound triple system, while also imparting enough momentum to the triple system to eject it from the cluster. However, that scenario may still require some fine tuning to retain a widely separated third body whose orbital velocity is much less than a typical globular cluster escape velocity. A remarkably similar system is PSR J1024-0719\cite{Kaplan2016}, which also has a low-mass low-metallicity widely separated companion star, and was suggested to have been ejected from a globular cluster. 

If ZTF J1406+1222 formed in the Galactic field, perhaps the most appealing solution is that the neutron star in ZTF J1406+1222 was formed by AIC rather than a core-collapse explosion. In this scenario, the neutron star was formed from a white dwarf that grew to larger than the Chandrasekhar mass $M_{\rm Ch} \sim 1.4 \, \rm M_\odot$ after accreting from the binary companion. In this case, the decrease in gravitational mass of the neutron star upon AIC is $\sim \! 0.1 \, \rm M_\odot$, resulting in a kick to the centre of motion of the binary of a few ${\rm km~s^{-1}}$ (or less for $M_2 < 1 \, \rm M_\odot$), small enough that the tertiary is more likely to remain bound. The neutron star could become a millisecond pulsar immediately (depending on the spin rate of the accreting white dwarf) or after further mass accretion from the binary companion, creating the spider system that we observe today. More sophisticated modelling of the possibilities outlined above should be performed in order to understand the formation of ZTF J1406+1222.

\end{methods}
\section{Data Availability}
Reduced HiPERCAM photometric data and LRIS spectroscopic data are availible at \\ https://github.com/kburdge/ZTFJ1406-1222. The X-ray observations already in the public domain, and their observation IDs have been supplied in the text. The ZTF data is also in the public domain. The proprietary period for the spectroscopic data will expire at the start of 2022, at which point the raw spectroscopic images will also be accessible via the Keck observatory archive. 

\section{Code Availability}

Upon request, the first author will provide code (primarily in python) used to analyze the observations, and any data used to generate figures (MATLAB was used to generate most of the figures). 




\end{document}